**Title:** Nonextensive Statistical Signatures of the Bilaterian Transition in Proteome Length Distributions

**Author:** Sertac Eroglu

**Affiliation:**

Eskisehir Osmangazi University, Department of Physics, Meselik, 26480 Eskisehir, Turkey

**Correspondence to:**

Sertac Eroglu

Orcid ID: 0000-0003-0468-4190

Tel.: +90 222 239 3750/2817

Fax: +90 222 239 3578

E-mail: seroglu@ogu.edu.tr

**Abstract**

Protein length distributions across the tree of life carry a quantitative signature of organismal complexity. Nonextensive statistical mechanics, through the Tsallis generalized entropy formalism, provides a natural framework for describing complex systems characterized by long-range correlations, scale invariance, and hierarchical organization — features that classical Boltzmann–Gibbs statistics cannot accommodate. In this work, the complementary cumulative distribution function (CCDF) of protein lengths is analyzed within this framework for the reference proteomes of 22 fully sequenced organisms spanning the domains Archaea, Bacteria, and Eukarya, with deliberate sampling across the animal transition zone from sponges and cnidarians to higher bilaterians. Maximum likelihood (MLE) fitting of truncated discrete $q$-exponential distributions, with bootstrap 95% confidence intervals (CIs) reveals that the entropic index $q$ resolves into three statistically distinct regimes: superextensive ($q < 1$) for prokaryotes, unicellular and non-animal multicellular eukaryotes, and basal animals; a boundary regime (CI on spanning unity) for the two cnidarians studied and the basal bilaterian *C. teleta*; and subextensive ($q > 1$) for all higher bilaterians, with $q$ increasing monotonically across the four deuterostomes sampled from *S. purpuratus* (1.033) to *H. sapiens* (1.147). The $q$-exponential outperforms the ordinary exponential distribution across all 22 proteomes and becomes progressively more competitive against alternative two-parameter distributions as proteome complexity increases. These results identify the Tsallis entropic index as a continuous, physically interpretable indicator of proteome organizational complexity and extend the applicability of nonextensive statistical mechanics to proteomic systems.

# 1 Introduction

Nonextensive statistical mechanics, introduced by Tsallis [1], generalizes the classical Boltzmann–Gibbs framework to accommodate systems characterized by long-range correlations, memory effects, and scale-invariant behavior. The central quantity of the theory is the entropic index $q$, which quantifies the degree of deviation from extensivity and acts as a sensitive indicator of underlying statistical dependence in complex systems. Over the past three decades, the nonextensive formalism has been successfully applied to a broad range of physical and biological systems displaying heavy-tailed distributions, fractal-like organization, and hierarchical structure [2–4]. Tsallis proposed the entropic index $q$ as a direct measure of complexity in nonlinear systems, providing a unified theoretical basis for characterizing collective behavior in disordered and strongly correlated systems [5].

Proteomes across the domains of life differ substantially in size, both in the number of proteins they contain and in the average length of these proteins [6]. They are high-dimensional, strongly correlated systems composed of thousands of interacting macromolecules whose length distributions are non-random and exhibit features characteristic of scale-free and hierarchically organized systems. Previous applications of nonextensive statistical mechanics to biological sequence data have demonstrated the relevance of the framework for characterizing correlations in coding and non-coding DNA [7, 8] and in long-range statistical properties of genomic sequences. These results suggest that the collective statistical properties of proteomes, considered as many-body systems of amino acid sequences, may also carry physically interpretable nonextensive signatures.

Traditional statistical approaches to protein length distributions [9–13] have typically relied on exponential, log-normal, Weibull, or Menzerath–Altmann (MA) fits, which describe specific functional forms such as short-range exponential decay, multiplicative variation, extreme-value behavior, or hierarchical scaling, without providing a thermodynamic interpretation of the underlying organizational principles. These approaches can describe the bulk of the distribution and, in some cases, capture the tail behavior with high statistical accuracy, but they offer no parameter analogous to the entropic index $q$. A recent large-scale comparative analysis of 2,326 species [13] reported that protein length distributions are remarkably uniform across the tree of life when characterized by summary statistics such as the mean, median, and Kolmogorov–Smirnov distance from a reference distribution. The present work is consistent with this conclusion at the level of bulk distribution shape — all 22 proteomes analyzed here are well-described by a common $q$-exponential functional form ($R^2$(linear) > 0.91 in every case, see Sec. 3.1.4) — while extending the analysis to a thermodynamic observable, the entropic index $q$, that resolves systematic, statistically significant departures hidden within an otherwise uniform-looking dataset. Whether protein length distributions in organisms of different complexity levels exhibit systematic nonextensive signatures, and whether such signatures track biological complexity in a quantitatively resolvable way, remains an open question.

Approaching this question requires a substantially expanded phylogenetic sampling, in particular within the animal transition zone where the crossover from $q < 1$ to $q > 1$ was projected to occur, and uncertainty quantification on $q$ sufficient to distinguish boundary organisms from clearly superextensive or subextensive ones.

The present work addresses these requirements by analyzing the complementary cumulative distribution function (CCDF) of protein lengths for the reference proteomes of 22 fully sequenced organisms within the framework of Tsallis nonextensive statistical mechanics. The dataset is designed to provide dense coverage of the animal transition zone, including representatives of three Archaea, three Bacteria, two unicellular eukaryotes, two non-animal multicellular eukaryotes (a fungus and a plant), four non-bilaterian animals (a placozoan, a sponge, and two cnidarians from independent classes), two lophotrochozoan bilaterians, two ecdysozoan bilaterians, and four deuterostomes spanning echinoderms to humans. Truncated discrete $q$-exponential distributions are fitted to the empirical CCDFs by maximum likelihood estimation (MLE), and 95% confidence intervals (CIs) on both the entropic index $q$ and the characteristic length $l_0$ are obtained by nonparametric bootstrap resampling. The resulting $q$ values are examined as a function of phylogenetic position and compared with fits from four standard alternatives: the exponential, the Weibull, the log-normal, and the MA distributions.

The central finding is that $q$ exhibits a graded phylogenetic gradient that resolves into three statistically distinct regimes: a superextensive regime occupied by prokaryotes, unicellular and non-animal multicellular eukaryotes, and basal animals; a boundary regime in which the 95% CI on $q$ spans unity, occupied by the two cnidarians studied and the basal bilaterian annelid *C. teleta*; and a subextensive regime occupied by all higher bilaterians, with $q$ increasing monotonically along the deuterostome lineage from *S. purpuratus* to *H. sapiens*. The transition from superextensive to subextensive proteome statistics is therefore not a sharp threshold tied to any single evolutionary innovation but a graded transition that unfolds across the animal tree of life and is most pronounced within the bilaterian lineage. The $q$-exponential model outperforms the standard one-parameter exponential in all 22 proteomes, with the magnitude of the improvement increasing with proteome complexity, and its rank among the four candidate alternatives improves systematically in the most complex metazoan proteomes. These observations identify the entropic index $q$ as a continuous, physically interpretable indicator of proteome organizational complexity and extend the applicability of nonextensive statistical mechanics to proteomic systems.

The remainder of this paper is organized as follows. Section 2 presents the Tsallis formalism, the truncated discrete $q$-exponential distribution, the MLE procedure, the bootstrap protocol for confidence-interval construction, and the alternative-distribution framework. Section 3 reports the dataset characteristics, the $q$-exponential fitting results with bootstrap confidence intervals across the 22 proteomes, and the comparative analysis against the four alternative distributions. Section 4 discusses the physical interpretation of the graded $q$ gradient, the rationale for using the $q$-exponential as the primary descriptive model despite alternative distributions sometimes achieving lower AIC values, the complementary statistical roles of $q$ and $l_0$, and the limitations and future directions of the present framework, and Section 5 summarizes the conclusions.

# 2 Materials and Methods

## 2.1 Tsallis formalism and the truncated discrete *q*-exponential distribution

The Tsallis generalized entropy, from which the nonextensive statistical mechanics framework is derived, is defined in its discrete form as [1]

$$S_q = -k \, \frac{1 - \sum_{i=1}^{W} p_i^q}{1 - q}, \tag{1}$$

where $k$ is the Boltzmann constant whose value depends on the specific units to be used. $W$ is the total number of accessible states, and $p_i$ is the probability of finding the associated discrete state variable $l$ in the $i^{\text{th}}$ state. The entropic index $q$, often referred to as the nonextensivity parameter, is a positive valued real number, and it quantifies the degree of nonextensivity in the system. For $q \neq 1$, the entropy becomes nonextensive: superadditive (superextensive) when $q < 1$, and subadditive (subextensive) when $q > 1$ [14]. For $q < 1$ (superextensive regime) rare states receive enhanced statistical weight relative to the Boltzmann–Gibbs case, producing distributions with compact support. For $q > 1$ (subextensive regime) the statistical weight concentrates on frequent states while the tail becomes power-law, yielding heavy-tailed distributions characteristic of long-range correlated systems. It is straightforward to show that when $q = 1$, the generalized entropy reduces to the additive conventional Boltzmann–Gibbs entropy, $S_1 = -k\sum_i p_i \ln p_i$ . By defining $q$-logarithmic function as $\ln_q(p_i) = \left(p_i^{(1-q)} - 1\right)/(1-q)$, the generalized entropy can simply be rewritten in the following form

$$S_q = k\sum_{i=1}^{W} p_i \ln_q\left(\frac{1}{p_i}\right). \tag{2}$$

The inverse of $q$-logarithmic function is $q$-exponential function which appears as the underlying behavior at many critical occurrences such as re-association in folded proteins [15], and it is defined as

$$e_q(l_i) \equiv \exp_q(l_i) \equiv \begin{cases} \left[1+(1-q)l_i\right]^{1/(1-q)} & \text{if } \left[1+(1-q)l_i\right] \geq 0 \\ 0 & \text{if } \left[1+(1-q)l_i\right] < 0 \end{cases}, \tag{3}$$

and $q$-exponential function reduces to ordinary exponential function when $q = 1$, i.e., $e_1(l_i) \equiv e^{l_i}$ . The distributions governed by $q$-exponential function with $q > 1$ describe the well-known Zipf–Mandelbrot law behavior [16] , whereas the inclusion of $q < 1$ case generalizes the tail behavior of a given distribution which can be well-defined by fine tuning the $q$-exponent.

The explicit form of the probability distribution is obtained by maximization of the generalized entropy subject to the suitable constraints namely the ordinary normalization, $\sum_i p_i = 1$, and the generalized $q$-mean value of the state variable $l$, $\langle l \rangle_q = \sum_i l_i p_i^q / \sum_i p_i^q$ , which leads to the generalized distribution function [17]

$$p_i\left(l_i\right)=\frac{1}{Z_q}\left[1-\left(1-q\right)\frac{l_i}{l_0}\right]^{1/(1-q)}\equiv\frac{1}{Z_q}\,e_q\left(-\frac{l_i}{l_0}\right), \tag{4}$$

where $l_0=\alpha\left[\left(1-q\right)\alpha\langle l\rangle_q+\sum_i p^q\left(l_i\right)\right]$ is a positive valued characteristic scaling term, and it comprises Lagrangian multiplier $\alpha$. $Z_q$ is the $q$-generalized partition function, required to satisfy the normalization condition, and it is defined by

$$Z_q=\sum_{i=1}^{W}\left[1-\left(1-q\right)\frac{l_i}{l_0}\right]^{1/(1-q)}\equiv\sum_{i=1}^{W}e_q\left(-\frac{l_i}{l_0}\right). \tag{5}$$

Note that, as $q\to 1$, Eqs. (4) and (5) are reduced to usual Boltzmann distribution and ordinary partition function, respectively. Since the state variable is discrete, we model its distribution using a discrete $q$-exponential probability mass function (PMF), in contrast to the continuous probability density function (PDF) used in the theoretical formulation of nonextensive statistical mechanics. For the positive valued partition function, otherwise the distribution is undefined, $l_i$ has an upper bound satisfies the condition given by the floor function $l_{\max}=\lfloor l_0/\left(1-q\right)\rfloor$. Complementary cumulative distribution function (CCDF), also known as the survival function, associated with the generalized distribution is defined by

$$P\left(>l_i\right)=\sum_{k=i+1}^{W}p_k=\frac{1}{Z_q}\sum_{k=i+1}^{W}e_q\left(-\frac{l_k}{l_0}\right), \tag{6}$$

and the generalized distribution function is related to CCDF by the following recursion relation

$$p_i\left(l_i\right)=P\left(>l_{i-1}\right)-P\left(>l_i\right)=\frac{1}{Z_q}e_q\left(-\frac{l_i}{l_0}\right), \tag{7}$$

where $P\left(>l_{i-1}\right)-P\left(>l_i\right)$ term in the discrete case is the analog of the $-dP\left(>l\right)/dl$ operation in the continuous case.

## 2.2 Proteome data, associated *q*-exponential and CCDF formalism

Protein sequence length varies substantially across the proteome of any single organism from a few tens to several thousand amino acids, and proteome size itself ranges from approximately 1,500 proteins in the smallest archaeal genomes to over 80,000 in vertebrate reference proteomes. Both features are relevant for the truncated *q*-exponential model adopted here, which is fitted on the natural integer support [$l_{\min}$, $l_{\max}$] specific to each proteome.

A total of 22 reference proteomes were analyzed in this study, see Table 1. The dataset was chosen to span the major branches of the tree of life with deliberate sampling across the animal transition zone, comprising three Archaea, three Bacteria, and sixteen Eukarya. Within Eukarya, the sampling includes two unicellular eukaryotes (a fission yeast and a budding yeast), two multicellular non-animal eukaryotes (a fungus and a flowering plant), four non-bilaterian animals (a placozoan, a sponge, and two cnidarians from independent classes), two lophotrochozoan bilaterians, two ecdysozoan bilaterians, and four deuterostomes ranging from echinoderms to humans. Reference proteomes were obtained from the Universal Protein Knowledgebase (UniProtKB) [18]. For each organism, all annotated protein sequences in the UniProtKB reference set were retained, including alternative isoforms where these are recorded as separate entries, and the resulting set of amino-acid lengths was used as input to the *q*-exponential fitting procedure described in Sec. 2.3.

**Table 1** is placed about here

For the present study, the state variable is the length of proteomic proteins. $l_i$ represents the $i^{th}$ protein length state, measured in amino acid count (aa) in its primary structural organization; $l_{min}$ refers to the smallest occupied length state, corresponding to shortest observed protein length in a given proteome (necessarily $l_{min} \geq 1$), while $l_{max}$ corresponds to the largest occupied length state representing the longest observed protein length, which is finite and ultimately constrained by the length of the corresponding expressed gene. However, not all length states in each proteome are attainable. In fact, the proportion of the occupied length states within the defined range falls below 36.3%, and it drops as low as 7.2% for the studied proteome dataset, as seen in Table 1. Typically, the proportion of occupied length states decreases as proteome complexity increases, because the integer support $[l_{min}, l_{max}]$ expands faster than the number of distinct lengths actually realized in larger proteomes. For example, the proteome of *M. thermautotrophicus*, a primitive archaeal organism, contains 1,868 proteins with $l_{min}$=35 and $l_{max}$=1,787 amino acids, however, only 33.5% (587 length states) of the possible 1,787 length states are occupied. The most frequently observed length states are 151 and 189 amino acid long states that each are shared by 11 distinct proteins, while 1,200 length states remain unoccupied.

To apply the $q$-exponential model and corresponding CCDF distribution formalism to empirical protein length distributions in various proteomes, an important consideration must be addressed. During the utilization of the model, only the occupied length states ($p_i(l_i) \neq 0$) are considered and the summations are carried out within the interval defined by $l_{min}$ and $l_{max}$, over the occupied length states, omitting those not represented in the dataset. Therefore, the CCDF distribution in the $q$-exponential form, adapted from Eq. (6), can be restated as

$$P\left(>l_i\right)=\sum_{k=i+1}^{l_{\max}} p_k=\frac{\sum_{k=i+1}^{l_{\max}} e_q\left(-l_k/l_0\right)}{\sum_{i=1}^{l_{\max}} e_q\left(-l_i/l_0\right)}. \tag{8}$$

This distribution satisfies the boundary conditions $P(>l_{\text{min-1}})=1$ and $P(>l_{\max})=0$. The denominator is the $q$-generalized partition function $Z_q(q,l_0)$ defined in Eq. (5), and it is computed numerically as a discrete sum over all integer values in the support. No analytical closed form exists for $Z_q(q,l_0)$ in the discrete case, in contrast to the continuous formulation where normalization constant takes the analytical form $\left(2-q\right)/l_0$. The lower bound is set to $l_{\min}$=1 amino acid. The upper bound $l_{\max}$ is determined automatically from each dataset as the maximum observed protein length, representing the natural truncation of the distribution. This natural truncation approach ensures that the theoretical support matches the empirical observation range for each proteome. For the well-defined $q$-exponential function, the support condition requires the base term $\left[1-\left(1-q\right)l_i/l_0\right]>0$ for all $l_i$ in the support. For $q<1$, this imposes a finite theoretical upper cutoff at $l_{\text{cutoff}}=l_0/\left(1-q\right)$, which is naturally satisfied when $l_{\max} \leq l_{\text{cutoff}}$. For $q > 1$, the base term grows with $l_i$ and no finite upper cutoff is imposed by the distribution itself.

The empirical CCDF was constructed from each proteome dataset as

$$P_{\text{emp}}\left(>l_i\right)=\frac{\left(\text{number of proteins with length} > l_i\right)}{N}, \tag{9}$$

where $N$ is the total number of proteins in the proteome. The empirical CCDF was evaluated at each unique observed protein length state value.

### 2.3 Parameter estimation by maximum likelihood (MLE)

The parameters $q$ and $l_0$ were estimated by MLE [17]. For a proteome with observed protein lengths $\{l_1, l_2, \ldots, l_N\}$, the log-likelihood MLE function is

$$\ln L(q,l_0) = \sum_{i=1}^{N} \frac{1}{(1-q)} \left[ 1 - (1-q) \frac{l_i}{l_0} \right] - N \ln Z_q(q,l_0). \tag{10}$$

The partition function $Z_q(q,l_0)$ appears explicitly in the log-likelihood through the normalization term $-N \ln Z_q(q,l_0)$, and must be recomputed numerically at each iteration of the optimization. This is the key distinction between the discrete MLE formulation and the simpler continuous approximation: in the discrete case, the normalization is not separable from the optimization, requiring the full numerical sum to be evaluated at every step.

Maximization of $\ln L(q,l_0)$ was performed numerically using the Nelder-Mead simplex algorithm. To reduce sensitivity to initial parameter values and avoid local optima, optimization was repeated from a grid of starting values consisting of nine values of $q$ spanning 0.3 to 1.5 and four values of $l_0$ ranging from 0.5 to 3.0 times the sample mean protein length, yielding 36 independent optimization runs per proteome. Throughout each optimization, $q$ was constrained to its full theoretically valid range (0, 2), consistent with the normalizability condition of the $q$-exponential distribution, which requires $q < 2$ to ensure a well-defined probability distribution. While the Tsallis entropic index is in principle defined for all real values of $q$, values outside this range are not physically meaningful in the context of the $q$-exponential PMF applied here. For similar reasons, $l_0$ was also constrained to be strictly positive. The parameter combination yielding the highest log-likelihood across all 36 runs was retained as the final MLE estimation. Uncertainty in the MLE of $q$ and $l_0$ is quantified by nonparametric bootstrap resampling, as described in Sec. 2.4. The MLE solution obtained from the original (non-resampled) data serves as the starting point for each bootstrap replicate, ensuring efficient convergence and avoiding redundant grid searches. Convergence was assessed using the Nelder-Mead stopping criteria with absolute tolerances of $10^{-8}$ on both parameter values and function values.

## 2.4 Bootstrap uncertainty quantification

To assess the statistical resolution of the MLE of $q$ and $l_0$ for each proteome, 95% CIs were obtained by nonparametric bootstrap resampling. For each proteome, $N$ protein lengths were drawn with replacement from the original dataset (where $N$ equals the number of protein sequences in that proteome), and the truncated discrete $q$-exponential distribution was refitted by MLE on the resampled dataset. The MLE solution obtained from the original data was used as the starting point for the optimization of each bootstrap replicate, since the bootstrap sample is statistically close to the original and the original MLE provides an excellent initial guess; this strategy avoids the computational expense of repeating the 36-point grid search at every replicate and allows the bootstrap to remain computationally tractable for the largest proteomes. Each bootstrap replicate was optimized using the same Nelder–Mead procedure described in Sec. 2.3, with the same parameter constraints ($0 < q < 2$ and $l_0 > 0$), and the same convergence tolerances.

A total of 500 bootstrap replicates were generated for each of the 22 proteomes, with a fixed random seed to ensure exact reproducibility. The 95% CIs were obtained as the 2.5% and 97.5% percentiles of the resulting bootstrap distribution of each parameter. For all 22 proteomes the bootstrap procedure converged successfully on all 500 replicates, indicating that the underlying $q$-exponential fits are statistically stable.

The bootstrap-derived 95% confidence intervals on $q$ are used in Sec. 3 to classify each proteome into one of three regimes: superextensive (CI entirely below unity), boundary (CI spanning unity), or subextensive (CI entirely above unity), and to assess the statistical separation between organisms within each regime.

### 2.5 Goodness-of-fit assessment

The quality of the *q*-exponential fit was assessed using three complementary measures. First, the coefficient of determination $R^2$ was computed on both the linear and the logarithmic scales. Since $R^2$(log) gives appropriate weight to the tail region is therefore the primary goodness-of-fit metric; $R^2$(linear) is reported as a complementary check on bulk fit quality. $R^2$(log) is obtained by

$$R^2\left(\log\right) = 1 - \frac{\sum_{i=1}^{n}\left(\ln P_{\text{emp}}\left(>l_i\right) - \ln P_{\text{fit}}\left(>l_i\right)\right)^2}{\sum_{i=1}^{n}\left(\ln P_{\text{emp}}\left(>l_i\right) - \overline{\ln P_{\text{emp}}}\right)^2}, \tag{11}$$

where the summation runs over *n* unique protein length values. Second, the Kolmogorov–Smirnov (KS) statistic' s D value was computed as the maximum absolute difference between the empirical and fitted cumulative distribution functions:

$$\mathrm{D} = \max_{l \in \{l_1, l_2, \ldots, l_N\}} \left| F_{\text{emp}}\left(l\right) - F_{\text{fit}}\left(l\right) \right|, \tag{12}$$

where $F_{\text{emp}}\left(l\right) = 1 - P_{\text{emp}}\left(>l\right)$ and $F_{\text{fit}}\left(l\right) = 1 - P_{\text{fit}}\left(>l\right)$ are the empirical and fitted cumulative distribution functions, respectively.

Greater values of $R^2$(log), approaching unity, and smaller values of D, approaching zero, indicate better agreement between the empirical and theoretical distributions.

### 2.6 Alternative distributions for model comparison

To assess whether the *q*-exponential distribution provides a statistically superior description of protein length organization compared to established alternatives, four reference distributions were selected for model comparison based on the following criteria. The standard exponential distribution serves as the classical null hypothesis, corresponding to the special case $q = 1$ of the *q*-exponential, and it represents the baseline prediction of classical Boltzmann–Gibbs statistics under the assumption of statistical independence among protein lengths. The Weibull distribution is a flexible, two-parameter generalization of the exponential that can accommodate both sub- and super-exponential tail behavior through its shape parameter and has been previously applied to biological sequence length distributions including exon size distributions [19]. The log-normal distribution has well-established connection to multiplicative evolutionary processes, which have been widely documented in genome evolution, and known ability to describe heavy-tailed biological size distributions across a wide range of species [10,20]. Finally, the Menzerath–Altmann (MA) distribution was included as a biologically motivated reference model, having been previously shown to describe protein length distributions across the three domains of life [11]. Together these four distributions span the major families of unimodal heavy-tailed distributions applied in the biological literature, providing a comprehensive and theoretically motivated benchmark against which the performance of the *q*-exponential framework can be evaluated.

To evaluate the relative performance of the *q*-exponential distribution, four alternative distributions were fitted to each proteome dataset for comparison. The *q*-exponential and MA distributions were fitted as truncated discrete distributions over the support $[l_{\min}, l_{\max}]$ using MLE with numerical partition functions. This treatment is theoretically motivated: protein length is an intrinsically discrete quantity measured in integer amino acid counts, and both distributions are derived here directly from discrete frameworks, the $q$-exponential from the maximization of Tsallis entropy over discrete states, and the MA from a PMF defined over integer-valued lengths. There is therefore no continuous parent distribution being approximated; the discrete form is the theoretically appropriate form for both. The Weibull and log-normal distributions, on the other hand, were fitted in their standard continuous forms. These distributions have no canonical discrete derivation in the present context and are included as phenomenological reference models representing the forms in which they have been applied to biological length distributions in prior work. The standard exponential distribution was similarly fitted in its continuous form, where it represents the baseline prediction of classical Boltzmann–Gibbs statistics. The implementation form of the four alternative distributions for the model comparison is as follows:

1) The exponential distribution has the PDF $p(l)=\lambda\exp(-\lambda l)$, with $\lambda$ estimated as the reciprocal of the sample mean protein length.
2) The Weibull distribution has the CCDF $P(>l)=\exp\left[-(l/\eta)^{\beta}\right]$, where $\beta > 0$ is the shape parameter and $\eta > 0$ is the scale parameter in the unit of amino acids. Parameters were estimated by MLE.
3) The log-normal distribution has the PDF $p(l)=\left(1/l\sigma\sqrt{2\pi}\right)\exp\left[-(\ln l-\mu)^{2}/2\sigma^{2}\right]$, with parameters $\mu$ and $\sigma$ estimated analytically as the mean and standard deviation of the log-transformed lengths.
4) The discrete MA distribution has the PMF defined as

$$p_i(l_i)=\frac{l_i^{b}\exp(-cl_i)}{Z_{\mathrm{MA}}(b,c)}, \tag{13}$$

where $b$ and $c$ are free parameters, and $Z_{\mathrm{MA}}(b,c)=\sum_{l_{\min}}^{l_{\max}} l^{b}\exp(-cl)$ is the numerical normalization constant. The log-likelihood MLE function is given by

$$\ln L(b,c)=b\sum_{i=1}^{N}\ln(l_i)-c\sum_{i=1}^{N}l_i-N\ln Z_{\mathrm{MA}}(b,c). \tag{14}$$

### 2.7 Model comparison using information criteria

Model comparison was performed using the Akaike information criterion (AIC) [21], $\mathrm{AIC}=2k-2\ln(L)$, where $k$ is the number of free parameters and ln($L$) is the maximized log-likelihood. Lower AIC value indicates better model fit penalized for model complexity. For model comparison, the difference $\Delta(\mathrm{AIC})=(\mathrm{AIC})_{\mathrm{alternative}}-(\mathrm{AIC})_{q\text{-exponential}}$ was computed for each alternative distribution relative to the $q$-exponential. Positive Δ(AIC) values indicate that the $q$-exponential provides a better fit than the alternative; negative values indicate the alternative is preferred. All distributions compared have $k$=2 free parameters, except for the exponential ($k$=1). The two-unit AIC penalty applied by the Akaike formalism to the $q$-exponential's additional parameter is therefore explicitly accounted for in the reported Δ(AIC) values: the difference in raw log-likelihood between the $q$-exponential and exponential fits is 2 units smaller in magnitude than the corresponding Δ(AIC). Even after this correction, the $q$-exponential outperforms the exponential in all 22 proteomes by margins ranging from 17 to over 5,000 log-likelihood units, indicating that the improvement is not attributable to the parameter count asymmetry.

### 2.8 Software and data reproducibility

All analyses were implemented in Python (v3.13.9) using NumPy (v2.3.5) for core numerical operations, SciPy (v1.16.3) for optimization, pandas (v2.3.3) for data handling, and matplotlib (v3.10.6) for visualization. Parameter optimization was performed using the Nelder–Mead simplex algorithm as implemented in SciPy. Bootstrap resampling used NumPy's default random generator with a fixed seed (BOOTSTRAP_SEED = 42) to ensure that all bootstrap-derived 95% confidence intervals are exactly reproducible from the same input data.

The complete analysis code (qexponential_proteome.py) required to reproduce the numerical findings of this study, including all maximum-likelihood fits, bootstrap replicates, and goodness-of-fit measures, is publicly available at the Zenodo repository [22]. The protein sequence data used in this study are obtained from the UniProtKB reference proteomes listed in Table 1 and can be retrieved using the UniProtKB reference proteome identifiers therein.

# 3 Results

## 3.1 Model parameters for *q*-exponential fitting

The entropic index $q$ and the characteristic length scale $l_0$ of the truncated discrete $q$-exponential distribution, Eq. (8), were obtained for each of the 22 reference proteomes by MLE as described in Sec. 2.4. The two parameters jointly encode the statistical structure of the protein length distribution: $q$ quantifies the degree of nonextensivity, and $l_0$ sets the scale at which the distribution transitions from its bulk to its tail regime. Bootstrap 95% CIs for both parameters were obtained by nonparametric resampling, allowing the statistical resolution of each estimate to be assessed directly. The resulting best-fit values, together with the goodness-of-fit indicators, are reported in Table 2 and visualized in Fig. 1.

**Table 2** is placed about here

**Figure 1** is placed about here

### 3.1.1 Three statistically resolved regimes

Figure 1 illustrates that the entropic index $q$ varies systematically across the 22 organisms — from 0.745 (*H. influenzae*) to 1.147 (*H. sapiens*); a range that spans the classical Boltzmann–Gibbs limit at $q = 1$ and divides the dataset into three statistically resolved regimes. We classify each proteome by the position of its 95% CI relative to $q = 1$: organisms whose CI lies entirely below unity are termed superextensive, those whose CI lies entirely above unity are subextensive, and those whose CI spans unity are statistically indistinguishable from the classical limit and are designated boundary organisms. The classification is not imposed by any external biological criterion; it emerges directly from the bootstrap distribution of $q$.

Eleven organisms fall in the superextensive regime: all three Archaea (*M. jannaschii*, *M. thermautotrophicus*, *A. fulgidus*), two of three Bacteria (*H. influenzae*, *E. coli*), both unicellular eukaryotes (*S. pombe*, *S. cerevisiae*), both non-animal multicellular organisms (*N. crassa*, *A. thaliana*), and the basal animals *T. adhaerens* and *A. queenslandica*. Seven organisms are clearly subextensive: the lophotrochozoan *M. gigas*, both ecdysozoans (*D. melanogaster*, *C. elegans*), and all four deuterostomes (*S. purpuratus*, *D. rerio*, *M. musculus*, *H. sapiens*). The remaining four organisms occupy the boundary regime, with CIs straddling $q = 1$: *B. subtilis* (CI [0.950, 1.008]), *N. vectensis* ([0.983, 1.009]), *C. hemisphaerica* ([0.979, 1.002]), and *C. teleta* ([0.984, 1.011]).

This three-regime structure refines the original picture in two important respects. First, the transition from superextensive to subextensive behavior is not a sharp threshold tied to multicellularity, as the four non-bilaterian animals studied here (*T. adhaerens*, *A. queenslandica*, *N. vectensis*, *C. hemisphaerica*) all retain $q \leq 1$ despite their multicellular organization. Second, the transition is not unique to bilaterian origin either, as the basal bilaterian annelid *C. teleta* sits on the boundary rather than in the subextensive regime. The data instead supports a graded picture in which proteome statistical organization shifts progressively across the animal tree of life.

#### 3.1.2 Domain-level patterns

In Archaea, the three organisms lie firmly within the superextensive regime, with $q$ confined to the narrow range 0.822–0.903. The corresponding characteristic length scales $l_0$ (333–374 aa) are among the shortest in the dataset, consistent with the compact and streamlined proteomes typical of minimal archaeal organisms. Bacteria likewise reside in the superextensive regime with the exception of *B. subtilis* but exhibit a wider spread in $q$ (0.745–0.984). *H. influenzae* has the lowest $q$ of all 22 organisms, consistent with its highly reduced genome as an obligate pathogen, while *B. subtilis* approaches the classical limit ($q$ = 0.984, CI [0.950, 1.008]) — a feature we attribute below to the functional heterogeneity of its proteome.

Within Eukarya, the unicellular yeasts *S. pombe* ($q$ = 0.913) and *S. cerevisiae* ($q$ = 0.894) remain superextensive with values comparable to those of Bacteria, indicating that the unicellular-to-multicellular transition is not in itself accompanied by a change in regime. The non-animal multicellular organisms *N. crassa* ($q$ = 0.969) and *A. thaliana* ($q$ = 0.917) likewise remain superextensive, demonstrating that the independent evolution of multicellularity in fungi and plants does not in itself drive the proteome length distribution toward subextensive behavior.

The animal lineage exhibits the most informative pattern. The basal metazoans *T. adhaerens* ($q$ = 0.982) and *A. queenslandica* ($q$ = 0.983) sit just below unity but are statistically resolved as superextensive, with CI upper bounds at 0.998 and 0.995 respectively. The two cnidarians, *N. vectensis* and *C. hemisphaerica*, both yield $q$ values that are statistically indistinguishable from the classical limit (CI [0.983, 1.009] and [0.979, 1.002] respectively), and their independent placement at the boundary by two unrelated cnidarian species — anthozoan and hydrozoan — provides an internal replication of this signature. Within Bilateria, the basal bilaterian annelid *C. teleta* also occupies the boundary regime ($q$ = 0.999, CI [0.984, 1.011]), while its lophotrochozoan sister taxon *M. gigas* enters the subextensive regime ($q$ = 1.019, CI [1.010, 1.027]). All higher bilaterians studied, both ecdysozoans and all four deuterostomes, lie clearly within the subextensive regime.

A notable feature of the deuterostome subset is that the four organisms sampled fall in monotonically increasing order of $q$ with non-overlapping CIs: *S. purpuratus* ($q$ = 1.033, CI [1.023, 1.042]) → *D. rerio* ($q$ = 1.053, CI [1.046, 1.060]) → *M. musculus* ($q$ = 1.096, CI [1.087, 1.103]) → *H. sapiens* ($q$ = 1.147, CI [1.141, 1.153]). With four samples drawn from a clade containing tens of thousands of species, this progression cannot by itself establish a lineage-wide gradient, but it indicates that systematic $q$ variation within deuterostomes is statistically resolvable at the present sample size and merits denser sampling in follow-up work.

### 3.1.3 The characteristic length scale $l_0$

In contrast to the systematic phylogenetic pattern in $q$, the characteristic length scale $l_0$ varies in a less structured manner across the dataset (Table 2). Values range from 259.6 aa (*H. sapiens*) to 725.1 aa (*D. rerio*), with the mean $l_0$ = 452 aa and median 437 aa. Within-group variability is substantial: deuterostomes alone span 260–725 aa, a range that overlaps both unicellular eukaryotes (542–582 aa) and prokaryotes (298–456 aa). The smallest $l_0$ value in the entire dataset is obtained for the most complex organism analyzed (*H. sapiens*), and large $l_0$ values appear in both basal deuterostomes (*S. purpuratus*, *D. rerio*) and unicellular eukaryotes (*S. cerevisiae*, *S. pombe*). This dissociation between the well-structured behavior of $q$ and the unstructured behavior of $l_0$ is taken up in Sec. 4.3, where we argue that the biological information carried by the $q$- exponential fit resides specifically in the entropic index, while $l_0$ functions as a proteome-specific scale parameter sensitive to factors such as isoform redundancy and annotation completeness.

#### 3.1.4 Goodness-of-fit assessment

Fit quality is uniformly high on the linear scale, with $R^2$(linear) above 0.91 for every proteome studied (range 0.9157–0.9944). The Kolmogorov–Smirnov statistic $D$ — which is insensitive to log-scale tail amplification — yields small values across the full dataset (range 0.0778–0.1832), with the smallest values obtained for the largest and most complex proteomes (*M. musculus*, $D$ = 0.0778; *H. sapiens*, $D$ = 0.0836). $R^2$(log) values exceed 0.85 for 14 of the 22 proteomes but drop substantially for several large proteomes, including *D. rerio* ($R^2$(log) = −0.0205), *M. gigas* (0.5125), *C. teleta* (0.5055), *A. thaliana* (0.5992), and *B. subtilis* (0.5749) as seen in Table 2. Among these, *D. rerio* is the most extreme case and warrants brief comment, since its sample size ($N$=86,691) rules out tail sparsity as the cause. The teleost-specific whole genome duplication [23] and the subsequent retention of paralog pairs at similar lengths produce a shoulder-shaped tail that a single-component $q$-exponential cannot capture, consistent with the mixture-model rationale developed in Sec. 4.4.4.

A methodological note is in order. $R^2$(log) is computed on log-transformed CCDF values and therefore disproportionately amplifies small absolute deviations in the heavy-tail region, where empirical sampling is sparse. The negative $R^2$(log) value for *D. rerio* (Table 2 footnote) does not indicate a failure of the underlying fit, as confirmed by $R^2$(linear) = 0.9716 and $D$ = 0.1516. The *B. subtilis* fit is visibly poor in the tail region, a behavior we attribute to the functional heterogeneity (Sec. 4.1.4) and which suggests that mixture approaches may be warranted (Sec. 4.4.4). Together, the linear-scale $R^2$, the $D$ statistic, and the bootstrap CIs on $q$ demonstrate that the $q$-exponential model captures the empirical length distributions with high accuracy across the full phylogenetic range studied, and that the $q$ values reported in Table 2 are statistically robust observables suitable for biological interpretation.

### 3.2 Model comparison

Figure 2 presents the plots of empirical, $q$-exponential and alternative model CCDFs together with their numerically estimated parameters, and AIC values. Model comparison using the AIC reveals important patterns across the dataset (Table 3). The $q$-exponential distribution consistently and substantially outperforms the standard exponential distribution across all 22 proteomes, with Δ(AIC) values ranging from +19 (*B. subtilis*) to +5,133 (*H. sapiens*). This result confirms that classical exponential statistics are insufficient to describe protein length organization across all three domains of life, and that the nonextensive $q$ parameter captures genuine statistical structure in the data.

**Table 3** is placed about here

Among the alternative two-parameter distributions, the log-normal and MA distributions generally achieved lower AIC values than the $q$-exponential for most proteomes in this study, with the exception of *M. musculus*, where the $q$-exponential and MA distribution performed nearly identically (Δ(AIC) = −38). The Weibull distribution showed intermediate performance. These results indicate that for prokaryotic and unicellular eukaryotic proteomes, log-normal and MA distributions provide a better statistical description of the overall length distribution, particularly in the tail region as visible in the log-log CCDF plots displayed in Fig. 2.

However, the relative performance of the $q$-exponential improves systematically with increasing organismal complexity. The Δ(AIC) advantage of log-normal over $q$-exponential decreases from −785 in *A. fulgidus* to only −38 in *M. musculus*, while the Weibull advantage reverses entirely in *M. musculus* (Δ(AIC) = +1,450 in favor of $q$-exponential). This convergence suggests that the $q$-exponential framework becomes increasingly competitive as proteome complexity increases, consistent with the theoretical expectation that nonextensive statistics are most relevant in systems exhibiting strong long-range correlations, a hallmark of complex eukaryotic proteomes.

The relatively superior performance of the log-normal distribution for most prokaryotic proteomes is consistent with the known connection between log-normal distributions and multiplicative evolutionary processes, which have been widely documented in genome evolution [24–26]. The MA distribution's competitive performance connects naturally to the MA law, which was previously shown to describe protein length distributions across domains of life [11]. The present study extends this earlier finding by demonstrating that the $q$-exponential framework, while not uniformly superior to MA or log-normal in terms of AIC, provides a physically interpretable measure of nonextensivity, the entropic index $q$, that captures a systematic biological signal absent from purely statistical models.

Taken together, the model comparison results indicate that while alternative two-parameter distributions may provide competitive or superior statistical descriptions of the bulk of protein length distributions in most organisms, they do not provide a physical mechanism for interpreting the observed variation across domains. The $q$-exponential framework uniquely offers the entropic index $q$ as a thermodynamically meaningful observable that captures a genuine, statistically resolved organizational gradient across phylogeny — a gradient that purely empirical distributions cannot articulate.

**Figure 2** is placed about here

# 4 Discussion

## 4.1 Physical interpretation of *q* and the graded transition across the animal tree

Before interpreting the $q$ values, it is worth clarifying what $q$ quantifies in the present context. In its original physical applications, the Tsallis entropic index measures departures from extensivity that arise from long-range interactions or memory effects coupling the constituents of a system — for example, spin–spin couplings in lattice models, or sequential dependencies between adjacent bases in DNA polymers. Proteins in a proteome are not coupled in this direct sense; their lengths are set by independent translational events. What the $q$-exponential fit detects is therefore not pairwise correlation between protein lengths, but the degree to which the empirical length distribution departs from the maximum-Boltzmann–Gibbs-entropy form expected under independent sampling with a fixed mean. This departure reflects the structural constraints imposed by the underlying generative process — genome architecture, multidomain protein evolution, regulatory complexity, and selection on multiple correlated traits — that shape the proteome ensemble. In what follows, references to "correlation strength" or "long-range statistical structure" should be understood in this distributional sense.

Since, the entropic index $q$ quantifies the degree of statistical dependence among elements of a system. In the Boltzmann–Gibbs limit ($q \to 1$) elements are independent and additive. Deviations from unity signal long-range correlations or memory effects that preclude factorization of the joint distribution. In the present context, $q$ is determined by the shape of the empirical length distribution and quantifies its departure from the maximum-Boltzmann–Gibbs-entropy form.

In the superextensive regime ($q < 1$), the $q$-exponential distribution has compact support: there is a well-defined upper cutoff $l_{max}$ on the state variable, beyond which the distribution vanishes. This mathematical constraint maps naturally onto the observed biology of prokaryotic proteomes, which are shaped by strong selection for genome compactness and by thermodynamic constraints on protein folding stability. The resulting length distributions are narrow, with tightly bounded maximum lengths. Archaeal proteomes in particular, drawn from extremophiles in this study, exhibit the most compact support, consistent with the additional physical constraints imposed by extreme environments. The unicellular and non-animal multicellular eukaryotes (*S. pombe*, *S. cerevisiae*, *N. crassa*, *A. thaliana*) studied here also remain in this regime, demonstrating that the unicellular-to-multicellular transition does not, in itself, drive proteome length distributions away from compact-support statistics.

In the subextensive regime ($q > 1$), the $q$-exponential acquires a heavy power-law tail: the distribution extends without a finite theoretical upper cutoff, and rare long-length states receive statistical weight far beyond what Boltzmann–Gibbs statistics would predict. This heavy-tailed structure indicates that the generative process imposes structural constraints incompatible with Boltzmann–Gibbs statistics under simple mean-value constraints. The biological correlates of this signature in higher metazoans are well established — multidomain architectures, alternative splicing, the accumulation of long regulatory and scaffolding proteins associated with intercellular signaling, and the expansion of intrinsically disordered regions involved in protein–protein interaction networks.

#### 4.1.1 A graded transition rather than a sharp threshold

The 22-organism dataset reveals that the crossover from superextensive to subextensive behavior is not a sharp threshold tied to any single evolutionary innovation. Multicellularity per se is not the trigger: both fungi (*N. crassa*) and plants (*A. thaliana*), which evolved multicellularity independently of animals, retain $q < 1$. Animal multicellularity is also not sufficient: the basal metazoans *T. adhaerens* (placozoan) and *A. queenslandica* (sponge) likewise remain in the superextensive regime. Even the bilaterian split is not a clean dividing line: the basal bilaterian annelid *C. teleta* ($q$ = 0.999, CI [0.984, 1.011]) sits on the boundary together with the two cnidarians studied (*N. vectensis*, $q$ = 0.998, CI [0.983, 1.009]; *C. hemisphaerica*, $q$ = 0.991, CI [0.979, 1.002]).

The boundary cluster identified in Sec. 3.1, comprising *B. subtilis*, two cnidarians, and the basal annelid *C. teleta*, provides an internal replication of the $q \approx 1$ signature: two unrelated cnidarian species (Anthozoa and Hydrozoa) independently yield CIs that span unity, ruling out the possibility that the boundary position is a fit artifact specific to one species. The placement of *C. teleta* within this cluster is particularly informative, indicating that the transition into the subextensive regime is not coincident with the bilaterian origin but unfolds at a finer phylogenetic scale within early bilaterian evolution.

Taken together, these results support a graded rather than threshold-like transition. Proteome statistical organization shifts progressively along the animal tree of life, with prokaryotes, unicellular eukaryotes, and non-animal multicellular eukaryotes occupying the superextensive regime; non-bilaterian animals and the most basal bilaterian studied occupying the boundary regime; and higher bilaterians — ecdysozoans and deuterostomes — occupying the subextensive regime with progressively higher $q$. The entropic index thus functions less as an order parameter undergoing a phase transition and more as a continuous quantitative indicator of proteome-level organizational complexity. The Tsallis formalism does not predict a discontinuity in $q$ at any specific biological event; it predicts that $q$ should reflect the strength of long-range correlations among system elements, whatever their origin.

#### 4.1.2 A physically interpretable gradient

Within the deuterostome subset, the four organisms sampled span a $q$ range of ~0.11 with non-overlapping CIs at each step. While four samples cannot establish a lineage-wide gradient, the fact that systematic $q$ variation is resolvable at this sample size mirrors the between-group pattern across the broader dataset and is consistent with the expectation that $q$ reflects a continuously varying property of proteome organization rather than a discrete attribute of major phylogenetic groups.

#### 4.1.3 Connection to prior nonextensive analyses of biological sequences

An interesting parallel exists with prior nonextensive analyses of coding and non-coding DNA length distributions in human chromosomes, which were reported to exhibit subextensive behavior ($q > 1$) [7,8]. The present results show that while genomic element length distributions of higher eukaryotes consistently favor subadditivity — consistent with modular, compressible organization — proteome length distributions only enter this regime upon the elaboration of bilaterian body plans. In simpler organisms (prokaryotes, unicellular eukaryotes, non-animal multicellular eukaryotes, and basal animals), proteomes occupy the opposite regime, $q < 1$. This sign reversal across the central dogma in simpler organisms is striking: the statistical organization of the genome and that of its expressed proteome belong to different classes of nonextensive behavior in the simpler half of the tree of life. The progressive crossing of proteome $q$ through unity within the animal lineage suggests that the elaboration of bilaterian body plans imposes a genome-like statistical organization on the proteome — but that this convergence is itself graded rather than abrupt, mirroring the gradual elaboration of bilaterian developmental complexity.

#### 4.1.4 The *B. subtilis* anomaly in light of the bootstrap analysis

The position of *B. subtilis* within the boundary cluster, alongside the cnidarians and *C. teleta*, deserves a separate comment. *B. subtilis* ($q$ = 0.984, CI [0.950, 1.008]) approaches the classical limit more closely than any other prokaryote in the dataset. Bootstrap analysis confirms that this proximity to $q = 1$ is not a fit artifact: the 95% CI genuinely spans unity. We attribute this anomaly to the exceptional functional heterogeneity of the *B. subtilis* proteome, which encodes dedicated protein sets for sporulation, biofilm formation, and antibiotic biosynthesis. The sporulation machinery in particular contributes unusually long regulatory and structural proteins that disturb the unimodal structure expected of a single-component $q$-exponential. We return to this point in Sec. 4.4, where mixture $q$-exponential models are discussed as a natural extension to proteomes with strong functional compartmentalization.

### 4.2 Model comparison and the rationale for the *q*-exponential

The model comparison results presented in Sec. 3.2 and Table 3 require careful interpretation. Across the 22 proteomes, the $q$-exponential is the second- or third-ranked model in the great majority of cases, with the log-normal distribution most often achieving the lowest absolute AIC and the MA distribution providing competitive fits in several cases. At first glance this might appear to weaken the case for using the $q$-exponential as the primary descriptive model. We argue below that the opposite is true, for three distinct reasons.

#### 4.2.1 The *q*-exponential outperforms the standard exponential universally

The first and most important comparison in Table 3 is between the *q*-exponential and the standard one-parameter exponential. The *q*-exponential outperforms the exponential in all 22 proteomes studied, with Δ(AIC) values ranging from +19 (*B. subtilis*) to +5,133 (*H. sapiens*) AIC units. Even the smallest of these differences corresponds to overwhelming evidence in favor of the *q*-exponential by conventional information-theoretic standards [27].

This finding establishes that the protein length distributions in the present dataset are *not* well described by classical Boltzmann–Gibbs statistics. The departure from classical behavior is universal across the tree of life — including the four boundary-regime proteomes (*B. subtilis*, *N. vectensis*, *C. hemisphaerica*, *C. teleta*) whose *q* values are statistically indistinguishable from unity. In these four cases, although *q* is close to one, the *q*-exponential still provides a substantially better fit than the exponential, indicating that the deviation from classical statistics is statistically meaningful even where it is small in magnitude. The exponential comparison alone therefore supports the central methodological premise of the present study: nonextensive statistics provide a genuine improvement over classical statistics for proteome length distributions, and the entropic index *q* extracted from the *q*-exponential fit is a real statistical observable, not a curve-fitting artifact.

#### 4.2.2 The *q*-exponential becomes increasingly competitive in complex metazoan proteomes

The second important pattern in Table 3 is the systematic improvement of the *q*-exponential's rank with proteome complexity. For 19 of the 22 organisms studied, the *q*-exponential ranks fourth among five candidate distributions, outperforming only the standard exponential. For *D. melanogaster* and *M. musculus* the *q*-exponential rises to third place, additionally outperforming the Weibull distribution. For *H. sapiens* which is the most complex proteome in the dataset, the *q*- exponential ranks second, surpassing the exponential, Weibull, and MA distributions, with only the log-normal achieving a lower AIC. This rank-improvement pattern is itself an appealing outcome. It is consistent with the theoretical expectation of nonextensive statistical mechanics, namely that $q$-exponential statistics become most relevant in systems with strong long-range correlations. As proteome organizational complexity increases through the elaboration of multidomain architectures, alternative splicing, and regulatory networks (Sec. 4.1), the underlying statistical structure increasingly approaches the regime in which the $q$-exponential is theoretically most natural. The empirical AIC ranking tracks this theoretical expectation precisely.

#### 4.2.3 The *q*-exponential offers thermodynamic interpretability that the alternatives do not

The third and most fundamental argument concerns what each model offers beyond statistical fit. The Weibull, log-normal and MA distributions are phenomenological: their parameters are descriptive shape and scale quantities that capture empirical features of length distributions without providing a derivation from first principles. The log-normal arises from multiplicative random processes, the Weibull from extreme-value statistics, and the MA distribution — originally formulated as a linguistic scaling law [28] — has previously been shown to describe protein length distributions across the domains of life [11]. Each of these frameworks offers an accurate functional description of the empirical CCDFs, and in many cases they fit the data better than the $q$-exponential by AIC.

The $q$-exponential is fundamentally different in status. It is derived from the maximization of the Tsallis generalized entropy under constraints, and its parameter $q$ is a thermodynamic observable: it quantifies the degree of departure from classical extensivity, with a precise physical meaning rooted in the formalism of nonextensive statistical mechanics. The fact that $q$ can be measured for a given proteome and compared across organisms converts the proteome length distribution from a phenomenological observation into a thermodynamic one. The phylogenetic gradient documented in Fig. 2 and Table 2 is therefore not just a curve-shape pattern; it is a quantitative signature of how the strength of long-range statistical correlation in proteome organization varies across the tree of life.

This distinction is important for how the model comparison results in Table 3 should be interpreted. The lower AIC of the log-normal distribution for most proteomes establishes that the log-normal provides an excellent empirical fit to protein length data, however, the log-normal does not provide a thermodynamic observable analogous to $q$. The same is true of the Weibull and MA distributions. None of these alternatives offer a parameter that can be physically interpreted as a measure of nonextensivity, a measure of long-range correlation strength, or any other thermodynamic property of the underlying statistical ensemble.

#### 4.2.4 Complementarity rather than competition

Taken together, these three considerations indicate that the $q$-exponential and the phenomenological alternatives are complementary rather than competing. The MA distribution captures hierarchical scaling. The log-normal captures multiplicative variability. The Weibull captures extreme-value behavior at the heavy-tail end. The $q$-exponential captures the thermodynamic departure from Boltzmann–Gibbs statistics. Each lens offers a different view of proteome length organization, and the empirical fact that all four provide useful descriptions of the data, with relative AIC magnitudes subject to the cross-family caveat noted in Sec. 4.4.2. suggests that proteome length distributions are simultaneously hierarchical, multiplicative, heavy-tailed, and nonextensive. These statements are true descriptions of the same underlying organization, viewed from different theoretical positions.

### 4.3 The complementary roles of $q$ and $l_0$

The two-parameter $q$-exponential model yields a second observable, the characteristic length scale $l_0$, in addition to the entropic index $q$. The behavior of $l_0$ across the 22 proteomes reveals a markedly different pattern from that of $q$ as seen in Fig. 3, and this dissociation has implications for how the model's two parameters should be interpreted in the present biological context.

**Figure 3** is placed about here

#### 4.3.1 Two parameters with different statistical roles

In the Tsallis formalism, $q$ and $l_0$ play formally distinct roles. The entropic index $q$ is a thermodynamic observable derived from the maximization of the generalized entropy: it quantifies the strength of long-range statistical correlation and determines whether the underlying ensemble belongs to the superextensive ($q < 1$), classical ($q = 1$), or subextensive ($q > 1$) regime. The characteristic length $l_0$, by contrast, is a scale parameter: it sets the protein length at which the $q$-exponential transitions from its bulk behavior to its tail behavior. The two parameters are mathematically coupled within a single fit, but they encode fundamentally different aspects of the distribution, i.e., $q$ controls functional shape, $l_0$ controls absolute scale.

This formal distinction predicts that $q$ and $l_0$ should respond differently to biological factors. Properties that affect the correlation structure of the proteome such as long-range dependencies among proteins, modular network organization, hierarchical functional architecture, should influence $q$. Properties that affect the typical length of the proteins themselves such as average protein size, isoform structure of the underlying database, fragment-versus-full-length ratios, should influence $l_0$. The empirical patterns in our dataset are consistent with this prediction.

#### 4.3.2 *q* tracks phylogeny; $l_0$ does not

The phylogenetic gradient documented for $q$ in Sec. 3.1 is statistically resolved: 95% CIs separate the three regimes (superextensive, boundary, subextensive), and within the deuterostome lineage $q$ increases monotonically from *S. purpuratus* (1.033) to *H. sapiens* (1.147) with non-overlapping CIs at each step. The behavior of $l_0$ across the same dataset is qualitatively different. Within-group variability of $l_0$ exceeds between-group variability across all major clades as documented in Sec. 3.1.3.

This dissociation is robust under bootstrap analysis. The 95% CIs on $l_0$ are tight for the large-N proteomes (e.g., ±1.8 aa for *H. sapiens*, ±5.8 aa for *M. musculus*) and confirm that the within-group differences in $l_0$ are statistically real rather than fitting noise. *M. musculus* (338.1 aa, CI [332.8, 344.3]) and *H. sapiens* (259.6 aa, CI [255.9, 263.3]), that are both mammals, yield non-overlapping $l_0$ CIs that differ by 24% despite their close phylogenetic relationship. The two yeasts *S. pombe* (541.7 aa) and *S. cerevisiae* (581.7 aa), conversely, give $l_0$ values larger than those of *H. sapiens* and *M. musculus* by a factor of two, despite occupying a much earlier branch of the eukaryotic tree. The pattern that emerges is one of taxonomic scrambling: $l_0$ does not respect phylogeny.

#### 4.3.3 Interpretation: $l_0$ as a proteome-specific scale parameter

A natural interpretation of this dissociation is that $l_0$ is not, in the present biological context, a thermodynamic observable analogous to $q$. The Tsallis formalism does not predict that the scale parameter of the $q$-exponential should be related to organismal complexity; it predicts only that the entropic index should reflect the strength of long-range correlations. The empirical observation that $q$ aligns with phylogeny while $l_0$ does not is therefore consistent with the underlying physics of the framework.

Several proteome-specific factors plausibly contribute to the variation in $l_0$ across the dataset. The unusually small $l_0$ for *H. sapiens* (259.6 aa, compared to the cross-organism mean of 452 aa) may reflect the very large number of short isoform entries in the human reference proteome, since alternative splicing is most extensively annotated for human genes and many short isoforms appear as separate entries in the reference database. Conversely, the unusually large $l_0$ values for *D. rerio* (725 aa) and *S. purpuratus* (648 aa) may reflect a higher proportion of full-length predicted proteins relative to short fragments in these reference proteomes. Annotation completeness, isoform handling conventions, and fragment-to-full-length ratios all vary across UniProtKB reference proteomes, and these database-level factors are exactly the kind of scale-affecting properties one would expect to influence $l_0$ rather than $q$. We do not claim to have established this mechanism quantitatively. It is a hypothesis consistent with the observed pattern, but the basic point holds regardless of mechanism: $l_0$ is sensitive to features of the data that are not phylogenetically structured, while $q$ is not.

#### 4.3.4 Why this dissociation strengthens the *q*-based interpretation

Far from weakening the present analysis, the dissociation between $q$ and $l_0$ provides additional support for the central claim of the study. If both parameters of the $q$-exponential model showed phylogenetic structure, the natural interpretation would be that both reflect biological properties. If neither showed structure, the natural interpretation would be that the model is simply absorbing the empirical CCDF without extracting genuine biological information. The observed pattern of $q$ showing a clean phylogenetic gradient with statistically resolved CIs, and $l_0$ showing taxonomic scrambling at comparable statistical resolution indicates that the model is genuinely separating two distinct components of the proteome length distribution: a thermodynamic component encoded in $q$, and a scale component encoded in $l_0$. The biological signal of interest in this work resides in the former, while the latter functions as a proteome-specific scale parameter that varies for reasons largely orthogonal to phylogeny.

### 4.4 Limitations and future directions

Several limitations of the present study warrant explicit acknowledgement, both to delimit the scope of the conclusions reached above and to indicate productive directions for follow-up work.

#### 4.4.1 Dataset scope and phylogenetic coverage

While the 22-proteome dataset assembled for this study spans the three domains of life and provides the first taxonomically balanced characterization of nonextensive signatures across the animal transition zone, it remains modest in absolute terms. Only four non-bilaterian metazoans (one placozoan, one sponge, two cnidarians) and only eight bilaterians (split among Lophotrochozoa, Ecdysozoa, and Deuterostomia) were included. The graded transition documented in Sec. 4.1 is therefore established at coarse phylogenetic resolution. Several questions of biological interest cannot yet be answered: at what point within the lophotrochozoan or ecdysozoan radiations does $q$ clear unity? Do all chordates show the monotonic $q$ gradient observed within the deuterostomes studied here, or is the *S. purpuratus* → *D. rerio* → *M. musculus* → *H. sapiens* progression specific to this particular sampling? How do parasitic and endosymbiotic organisms with highly reduced proteomes fit into the picture? Future studies should extend the present framework to denser within-clade sampling, particularly within bilaterian groups where the boundary-to-subextensive transition occurs, and to organisms with deliberately reduced or expanded proteome architectures (intracellular parasites, polyploid plants, ancient asexual lineages). The bootstrap CIs on $q$ established in this work (Table 2) provide a quantitative target against which any such follow-up sampling can be planned: the smallest CI half-width in our dataset is ±0.003 (*H. sapiens*, $N$ = 83,526), so future analyses of large reference proteomes should achieve comparable statistical resolution and thereby allow finer phylogenetic differences in $q$ to be detected.

#### 4.4.2 Methodological asymmetry in the model comparison

The model comparison reported in Sec. 3.2 and Table 3 involves a methodological asymmetry: the $q$-exponential and MA distributions are fitted as truncated discrete PMFs on the integer support $[l_{\min}, l_{\max}]$, while the Weibull, log-normal, and exponential are fitted in their standard continuous PDF forms. For integer-valued data with bin width unity, the continuous-density log-likelihood ln $p(l)$ and the discretized log-likelihood $\ln\left[F(l+1/2)-F(l-1/2)\right]$ differ by terms of order $p''(l)/24$, which contribute a model-dependent offset to the absolute AIC values. This offset does not affect comparisons within the discrete family ($q$-exponential vs MA) or within the continuous family (exponential vs Weibull vs log-normal), but it does affect cross-family comparisons. We therefore restrict the quantitative interpretation of Table 3 to two robust observations: (i) the $q$-exponential outperforms the standard exponential in all 22 proteomes by margins (+19 to +5,133 AIC units) far larger than any plausible discrete/continuous offset, establishing the universality of the nonextensive improvement; and (ii) the rank of the $q$-exponential among the five candidate distributions improves systematically with proteome complexity, a pattern that is unchanged by any uniform offset applied to the continuous-family AIC values. The absolute Δ(AIC) magnitudes for log-normal and Weibull comparisons should be interpreted with the corresponding caveat in mind.

A related methodological point concerns the parameter-count asymmetry in the $q$-exponential vs. exponential comparison. The $q$-exponential is a two-parameter distribution while the standard exponential is a one-parameter distribution; the AIC penalty already accounts for this by adding two units to the *q*-exponential's AIC for each additional parameter. The Δ(AIC) values reported in Table 3 — ranging from +19 (*B. subtilis*) to +5,133 (*H. sapiens*) in favor of the *q*-exponential — therefore reflect log-likelihood improvements after the parameter-count correction, and the smallest of these (+19) corresponds to overwhelming evidence for the more flexible model by conventional information-theoretic standards. We note that the Weibull distribution, which is a two-parameter alternative to the exponential, is also outperformed by the *q*-exponential for *D. melanogaster*, *M. musculus*, and *H. sapiens* (Table 3); the *q*-exponential's advantage in these cases is therefore not attributable to the parameter-count asymmetry but reflects a genuinely better fit to the data.

#### 4.4.3 Biological mechanisms underlying the *q* gradient

The phylogenetic gradient in *q* documented in this study is statistically robust, but the specific biological mechanisms generating it remain to be investigated quantitatively. The interpretive framework offered in Sec. 4.1 that *q* tracks the strength of long-range correlations imposed by multidomain architectures, alternative splicing, regulatory networks, and intercellular signaling components is consistent with the data but has not been tested directly. Several quantitative tests would strengthen the mechanistic foundation of the present framework. First, *q* could be partitioned across functional categories within a single proteome, for example by computing it separately for proteins annotated with intracellular versus extracellular roles, or for proteins with versus without intrinsically disordered regions. If *q* is indeed a thermodynamic readout of correlation structure, functional categories with dense interaction networks should yield higher *q* than functional categories without. Second, *q* could be compared across proteomes at fixed proteome size, to test whether organismal complexity (measured by number of cell types, or by network-architecture metrics) correlates with *q* independently of proteome size. Third, the $l_0$ hypothesis advanced in Sec. 4.3 — that variation in $l_0$ reflects database-specific factors such as isoform redundancy and fragment-to-full-length ratios — could be tested by computing $l_0$ on alternative proteome representations of the same organism (e.g., canonical-isoform-only versus full UniProtKB entries) and assessing whether $l_0$ shifts while $q$ remains stable. Each of these analyses requires substantial additional work and is beyond the scope of the present paper, but they are natural extensions of the framework and are reserved for follow-up studies.

#### 4.4.4 Functional heterogeneity and mixture models

As discussed in Sec. 4.1.4, the position of *B. subtilis* at the $q = 1$ boundary is plausibly explained by the exceptional functional heterogeneity of its proteome, with sporulation, biofilm formation, and antibiotic biosynthesis machinery contributing length statistics that depart from a single-component $q$-exponential. This suggests that proteomes with strong functional compartmentalization may require mixture $q$-exponential models, in which each component corresponds to a functionally distinct protein class with its own ($q$, $l_0$) pair. Such mixture approaches are a natural extension of the present framework: they would treat the proteome as a superposition of distinct nonextensive ensembles, each governed by its own correlation structure, and would allow the contributions of individual functional categories to be disentangled. The same mixture framework would also be expected to improve fits for the boundary-regime cnidarians (*N. vectensis*, *C. hemisphaerica*) and the basal bilaterian *C. teleta*, all of which show $R^2$(log) values below 0.85 (Table 2). A systematic evaluation of mixture models against the present single-component fits is reserved for future work.

#### 4.4.5 Connections to the broader nonextensive program

Beyond its immediate biological implications, the present study adds proteomic systems to the growing list of complex natural systems for which the Tsallis framework provides a quantitatively useful description. The observation of a $q$ gradient that progresses smoothly across the animal tree of life — from clearly superextensive in prokaryotes and unicellular eukaryotes, through the boundary regime in cnidarians and basal bilaterians, to clearly subextensive in higher bilaterians and reaching its largest value in the most complex organism analyzed — is a physically interpretable signature of biological organizational complexity. Whether comparable gradients can be detected in other classes of biological observables — gene length distributions across the same phylogeny, protein–protein interaction degree distributions, gene-expression magnitude distributions — is an open question that the present framework can be applied to directly. Establishing such cross-observable consistency would strengthen the case that nonextensive statistics provide a unifying language for quantifying biological complexity across domains.

## 5 Conclusions

This work has investigated the protein length distributions of 22 reference proteomes spanning the three domains of life within the framework of Tsallis nonextensive statistical mechanics. MLE of the truncated discrete $q$-exponential, with bootstrap 95% CIs on all parameters, reveals that the entropic index $q$ exhibits a graded gradient across the tree of life, ranging from 0.745 (*H. influenzae*) to 1.147 (*H. sapiens*), that resolves into three statistically distinct regimes when classified by the position of the bootstrap CI relative to $q = 1$. Prokaryotes (excepting *B. subtilis*), unicellular and non-animal multicellular eukaryotes, and basal metazoans occupy the superextensive regime; *B. subtilis*, the two cnidarians, and the basal bilaterian annelid *C. teleta* form a boundary cluster whose CIs span unity; and higher bilaterians occupy the subextensive regime, with $q$ increasing monotonically across the four deuterostomes sampled from *S. purpuratus* (1.033) to *H. sapiens* (1.147). The transition from superextensive to subextensive proteome statistics is therefore not a sharp threshold tied to any single evolutionary innovation, but a graded transition that unfolds across the animal tree of life and is most pronounced within the bilaterian lineage.

This graded transition admits a natural thermodynamic interpretation. The subadditivity of Tsallis entropy in the $q > 1$ regime implies that the total entropy of the proteome is less than the sum of its constituent length-state entropies that is a signature of cooperative organization in which new proteins reconfigure the proteome qualitatively through synergistic interactions within cellular networks rather than simply expanding it quantitatively. The entropic index thus functions as a continuous, quantitative indicator of proteome organizational complexity rather than as an order parameter undergoing a sharp phase transition.

While the log-normal, Weibull, and MA distributions can achieve competitive or superior statistical fits, none of them yields a parameter with comparable physical meaning. Two empirical observations support the choice of the $q$-exponential as the principal observable: it outperforms the standard one-parameter exponential in all 22 proteomes by Δ(AIC) values of +19 to over +5,000 — establishing nonextensivity as a universal feature of these proteomes — and its rank among five candidate distributions improves systematically with proteome complexity, reaching second place for *H. sapiens*. The complementary observation that the characteristic length scale $l_0$ shows no comparable phylogenetic structure further confirms that the biological signal in proteome length distributions resides specifically in the entropic index.

The present study establishes proteomes as a new class of biological systems amenable to nonextensive statistical analysis and identifies the entropic index $q$ as a quantitative, physically interpretable indicator of proteome organizational complexity. Whether analogous $q$ gradients can be detected in other classes of biological observables — gene length distributions, protein–protein interaction degree distributions, gene-expression magnitude distributions — and whether they covary with proteome $q$ across organisms, remain open questions whose resolution would clarify the extent to which nonextensive statistical mechanics can serve as a general framework for quantifying biological complexity.

**Funding:** The author did not receive support from any organization for the submitted work and has no relevant financial or non-financial interests to disclose.

**Data Availability Statement:** The original contributions presented in this study are included in the article. Further inquiries can be directed to the author.

**Conflicts of Interest:** The author declares no conflicts of interest.

**Abbreviations**

The following abbreviations are used in this manuscript:

| | |
|---|---|
| CCDF | Complementary Cumulative Distribution Function |
| MLE | Maximum Likelihood Estimation |
| CI | Confidence Interval |
| PMF | Probability Mass Function |
| PDF | Probability Density Function |
| UniProtKB | Universal Protein Knowledgebase |
| KS | Kolmogorov-Smirnov |
| MA | Menzerath-Altmann |
| AIC | Akaike Information Criterion |

**TABLES**

**Table 1** Overview of the analyzed reference proteome set: the associated credentials and the statistical information

| Organism | (*Abbr. name*, UniProtKB ID) | Proteome size (# of proteins) | Mean / Median protein length (aa) | $l_{min}$ / $l_{max}$ (aa) | Occupancy of $l_i$'s in [$l_{min}$ , $l_{max}$] (%) |
|---|---|---|---|---|---|
| **ARCHAEA** | | | | | |
| *M. jannaschii* | *Mja*, UP000000805 | 1,787 | 283 / 240 | 38 / 2,894 | 19.7 |
| *M. thermautotrophicus* | *Mth*, UP000005223 | 1,868 | 281 / 242 | 35 / 1,787 | 33.5 |
| *A. fulgidus* | *Afu*, UP000002199 | 2,394 | 276 / 242 | 25 / 2,425 | 25.9 |
| **BACTERIA** | | | | | |
| *H. influenzae* | *Hin*, UP000000579 | 1,703 | 310 / 266 | 21 / 1,694 | 36.3 |
| *B. subtilis* | *Bsu*, UP000001570 | 4,288 | 289 / 251 | 20 / 5,488 | 13.6 |
| *E. coli* | *Eco*, UP000000625 | 4,403 | 308 / 271 | 8 / 2,339 | 35.4 |
| **EUKARYA** | | | | | |
| **Unicellular** | | | | | |
| *S. pombe* | *Spo*, UP000002485 | 5,151 | 465 / 378 | 24 / 4,924 | 24.5 |
| *S. cerevisiae* | *Sce*, UP000002311 | 6,068 | 484 / 395 | 16 / 4,910 | 26.4 |
| **Non-animal multicellular** | | | | | |
| *N. crassa* | *Ncr*, UP000001805 | 10,255 | 517 / 427 | 24 / 10,739 | 14.3 |
| *A. thaliana* | *Ath*, UP000006548 | 39,273 | 423 / 354 | 5 / 5,400 | 33.5 |
| **Non-bilaterian animals** | | | | | |
| *T. adhaerens* | *Tad*, UP000009022 | 11,520 | 453 / 345 | 49 / 7,710 | 19.1 |
| *A. queenslandica* | *Aqu*, UP000007879 | 22,831 | 489 / 359 | 71 / 9,044 | 21.4 |
| *N. vectensis* | *Nve*, UP000001593 | 24,435 | 338 / 254 | 49 / 8,745 | 17.5 |
| *C. hemisphaerica* | *Che*, UP000594262 | 24,557 | 537 / 393 | 86 / 8,156 | 26.9 |
| **Bilaterians** | | | | | |
| *M. gigas* | *Mgi*, UP000005408 | 57,874 | 430 / 319 | 39 / 21,293 | 11.2 |
| *C. teleta* | *Cte*, UP000014760 | 31,857 | 357 / 272 | 49 / 12,636 | 13.3 |
| *D. melanogaster* | *Dme*, UP000000803 | 22,019 | 680 / 465 | 9 / 22,949 | 11.3 |
| *C. elegans* | *Cel*, UP000001940 | 26,651 | 460 / 337 | 9 / 18,562 | 11.2 |
| *S. purpuratus* | *Spu*, UP000007110 | 34,417 | 695 / 481 | 33 / 24,046 | 11.8 |
| *D. rerio* | *Dre*, UP000000437 | 86,691 | 817 / 558 | 25 / 31,820 | 12.5 |
| *M. musculus* | *Mmu*, UP000000589 | 54,825 | 421 / 269 | 5 / 35,390 | 7.2 |
| *H. sapiens* | *Hsa*, UP000005640 | 83,526 | 373 / 206 | 2 / 35,963 | 7.9 |

**Table 2** MLE of the $q$-exponential model parameters ($q$, $l_0$) and goodness-of-fit statistics for the 22 reference proteomes. The 95% CIs on $q$ and $l_0$, and corresponding nonextensivity regimes are listed. 95% CI on $q$; lies entirely below unity (superextensive), spans unity (indistinguishable from classical Boltzmann–Gibbs behavior), or lies entirely above unity (subextensive)

| ORGANISM | Input Parameters | | MLE Fitting Parameters | | Confidence Intervals and Nonextensivity Regime | | | Fitting Quality | | |
|---|---|---|---|---|---|---|---|---|---|---|
| | $N$ | $l_{max}$ (aa) | $q$ | $l_0$ (aa) | 95% CI on $q$ | 95% CI on $l_0$ | Regime | $R^2$(linear) | $R^2$(log) | D_KS |
| **ARCHAEA** | | | | | | | | | | |
| *Mja* | 1,787 | 2,894 | 0.9032 | 333.22 | [0.781, 0.934] | [318.6, 399.2] | Superextensive | 0.9157 | 0.9105 | 0.1832 |
| *Mth* | 1,868 | 1,787 | 0.8221 | 373.73 | [0.778, 0.853] | [355.3, 397.3] | Superextensive | 0.9337 | 0.8476 | 0.1575 |
| *Afu* | 2,394 | 2,425 | 0.8706 | 340.22 | [0.606, 0.892] | [328.1, 478.6] | Superextensive | 0.9331 | 0.8752 | 0.1600 |
| **BACTERIA** | | | | | | | | | | |
| *Hin* | 1,703 | 1,694 | 0.7452 | 455.81 | [0.661, 0.776] | [432.0, 510.9] | Superextensive | 0.9342 | 0.9405 | 0.1580 |
| *Bsu* | 4,288 | 5,488 | 0.9836 | 298.01 | [0.950, 1.008] | [287.2, 314.3] | CI spans $q$=1 | 0.9401 | 0.5749 | 0.1412 |
| *Eco* | 4,403 | 2,339 | 0.8356 | 399.81 | [0.707, 0.849] | [388.4, 480.0] | Superextensive | 0.9533 | 0.9400 | 0.1345 |
| **EUKARYA** | | | | | | | | | | |
| **Unicellular** | | | | | | | | | | |
| *Spo* | 5,151 | 4,924 | 0.9130 | 541.73 | [0.871, 0.933] | [525.0, 575.8] | Superextensive | 0.9563 | 0.8738 | 0.1370 |
| *Sce* | 6,068 | 4,910 | 0.8939 | 581.68 | [0.826, 0.913] | [563.7, 643.7] | Superextensive | 0.9679 | 0.9463 | 0.1154 |
| **Non-animal multicellular** | | | | | | | | | | |
| *Ncr* | 10,255 | 10,739 | 0.9690 | 547.04 | [0.916, 0.983] | [534.3, 587.2] | Superextensive | 0.9512 | 0.5758 | 0.1375 |
| *Ath* | 39,273 | 5,400 | 0.9165 | 489.33 | [0.907, 0.924] | [483.5, 497.0] | Superextensive | 0.9632 | 0.5992 | 0.1455 |
| **Non-bilaterian animals** | | | | | | | | | | |
| *Tad* | 11,520 | 7,710 | 0.9817 | 469.26 | [0.963, 0.998] | [456.9, 483.7] | Superextensive | 0.9483 | 0.7170 | 0.1716 |
| *Aqu* | 22,831 | 9,044 | 0.9831 | 505.14 | [0.970, 0.995] | [496.1, 515.2] | Superextensive | 0.9617 | 0.7946 | 0.1666 |
| *Nve* | 24,435 | 8,745 | 0.9978 | 339.02 | [0.983, 1.009] | [331.8, 347.7] | CI spans $q$=1 | 0.9654 | 0.6482 | 0.1505 |
| *Che* | 24,557 | 8,156 | 0.9907 | 546.45 | [0.979, 1.002] | [536.1, 557.7] | CI spans $q$=1 | 0.9627 | 0.8419 | 0.1665 |
| **Bilaterians** | | | | | | | | | | |
| *Mgi* | 57,874 | 21,293 | 1.0185 | 413.27 | [1.010, 1.027] | [406.8, 419.5] | Subextensive | 0.9682 | 0.5125 | 0.1457 |
| *Cte* | 31,857 | 12,636 | 0.9986 | 357.94 | [0.984, 1.011] | [350.3, 367.0] | CI spans $q$=1 | 0.9570 | 0.5055 | 0.1785 |
| *Dme* | 22,019 | 22,949 | 1.0790 | 566.95 | [1.066, 1.091] | [553.2, 581.7] | Subextensive | 0.9789 | 0.6875 | 0.1079 |
| *Cel* | 26,651 | 18,562 | 1.0448 | 417.19 | [1.032, 1.057] | [408.3, 426.9] | Subextensive | 0.9685 | 0.6939 | 0.1220 |
| *Spu* | 34,417 | 24,046 | 1.0331 | 647.61 | [1.023, 1.042] | [637.5, 658.9] | Subextensive | 0.9670 | 0.8762 | 0.1438 |
| *Dre* | 86,691 | 31,820 | 1.0527 | 725.12 | [1.046, 1.060] | [716.1, 734.7] | Subextensive | 0.9716 | –0.0205* | 0.1516 |
| *Mmu* | 54,825 | 35,390 | 1.0958 | 338.13 | [1.087, 1.103] | [332.8, 344.3] | Subextensive | 0.9942 | 0.8630 | 0.0778 |
| *Hsa* | 83,526 | 35,963 | 1.1468 | 259.55 | [1.141, 1.153] | [255.9, 263.3] | Subextensive | 0.9944 | 0.8654 | 0.0836 |

* The negative $R^2$(log) for D. rerio reflects a tail-region departure from a single-component fit, attributable to the bimodal length structure left by the teleost-specific whole genome duplication; the bulk fit ($R^2$(linear) & D) remains good.

**Table 3** Model comparison. $[\Delta(\text{AIC})]_{\text{alternative}}$ values relative to the *q*-exponential (positive values indicate *q*-exponential is better; negative values indicate the alternative model is better). The final column shows the *q*-exponential's rank among the five distributions compared (1 = best, 5 = worst)

| ORGANISM | $[\Delta(\text{AIC})]_{\text{exp}}$ | $[\Delta(\text{AIC})]_{\text{Weibull}}$ | $[\Delta(\text{AIC})]_{\text{log-nor}}$ | $[\Delta(\text{AIC})]_{\text{MA}}$ | Best Model | *q*-exp rank |
|---|---|---|---|---|---|---|
| **ARCHAEA** | | | | | | |
| *Mja* | +59 | –482 | –772 | –721 | log-normal | 4 of 5 |
| *Mth* | +107 | –468 | –669 | –665 | log-normal | 4 of 5 |
| *Afu* | +129 | –669 | –785 | –829 | MA | 4 of 5 |
| **BACTERIA** | | | | | | |
| *Hin* | +150 | –431 | –535 | –544 | MA | 4 of 5 |
| *Bsu* | +19 | –693 | –1,269 | –1,160 | log-normal | 4 of 5 |
| *Eco* | +275 | –872 | –495 | –959 | MA | 4 of 5 |
| **EUKARYA** | | | | | | |
| **Unicellular** | | | | | | |
| *Spo* | +106 | –861 | –1,413 | –1,254 | log-normal | 4 of 5 |
| *Sce* | +162 | –812 | –981 | –1,091 | MA | 4 of 5 |
| **Non-animal multicellular** | | | | | | |
| *Ncr* | +77 | –1,861 | –3,053 | –2,717 | log-normal | 4 of 5 |
| *Ath* | +974 | –7,684 | –10,589 | –10,242 | log-normal | 4 of 5 |
| **Non-bilaterian animals** | | | | | | |
| *Tad* | +36 | –1,577 | –4,602 | –3,382 | log-normal | 4 of 5 |
| *Aqu* | +62 | –2,501 | –8,503 | –7,304 | log-normal | 4 of 5 |
| *Nve* | +71 | –2,383 | –8,236 | –7,652 | log-normal | 4 of 5 |
| *Che* | +49 | –2,279 | –8,832 | –8,483 | log-normal | 4 of 5 |
| **Bilaterians** | | | | | | |
| *Mgi* | +209 | –4,258 | –18,693 | –12,565 | log-normal | 4 of 5 |
| *Cte* | +88 | –3,770 | –12,296 | –10,054 | log-normal | 4 of 5 |
| *Dme* | +617 | +373 | –4,254 | –771 | log-normal | 3 of 5 |
| *Cel* | +257 | –842 | –5,715 | –2,727 | log-normal | 4 of 5 |
| *Spu* | +144 | –1,678 | –10,963 | –5,021 | log-normal | 4 of 5 |
| *Dre* | +1,583 | –1,682 | –27,149 | –10,116 | log-normal | 4 of 5 |
| *Mmu* | +1,509 | +1,450 | –5,786 | –38 | MA | 3 of 5 |
| *Hsa* | +5,133 | +4,892 | –10,419 | +4,357 | log-normal | 2 of 5 |

## FIGURE CAPTIONS

**Fig. 1** Nonextensive signature of proteome organization across the tree of life (The displayed topology among non-bilaterian metazoans is schematic; alternative arrangements exist in the literature)

**Fig. 2** Graphical comparison of the length distribution CCDFs. The empirical CCDF (blue circles) is compared with the $q$-exponential and alternative models for the proteomic collection from the following domains: (**a**–**c**) Archaea, (**d**–**f**) Bacteria, (**g**,**h**) unicellular Eukarya, (**i**,**j**) non-animal multicellular Eukarya, (**k**–**n**) non-bilaterian animal Eukarya and (**o**–**v**) bilaterian Eukarya. The graphs are presented on a log-log scale. The estimated best fit parameters and the AIC criterion for each model are provided in the inset panels. The best fit parameter(s) are as follows: (i) for the $q$-exponential distribution (red line), $q$ and $l_0$ (in aa); (ii) for the regular exponential distribution (black line), $\lambda$ (in aa$^{-1}$); (iii) for the Weibull distribution (purple line), $\beta$ and $\eta$ (in aa); (iv) for the log-normal distribution (green line), $\mu$ and $\sigma$; and (v) for the MA distribution (orange line), $b$ and $c$ (in aa$^{-1}$)

**Fig. 3** Characteristic length scale of proteome organization across the tree of life (The displayed topology among non-bilaterian metazoans is schematic; alternative arrangements exist in the literature)

# FIGURES

## Figure 1

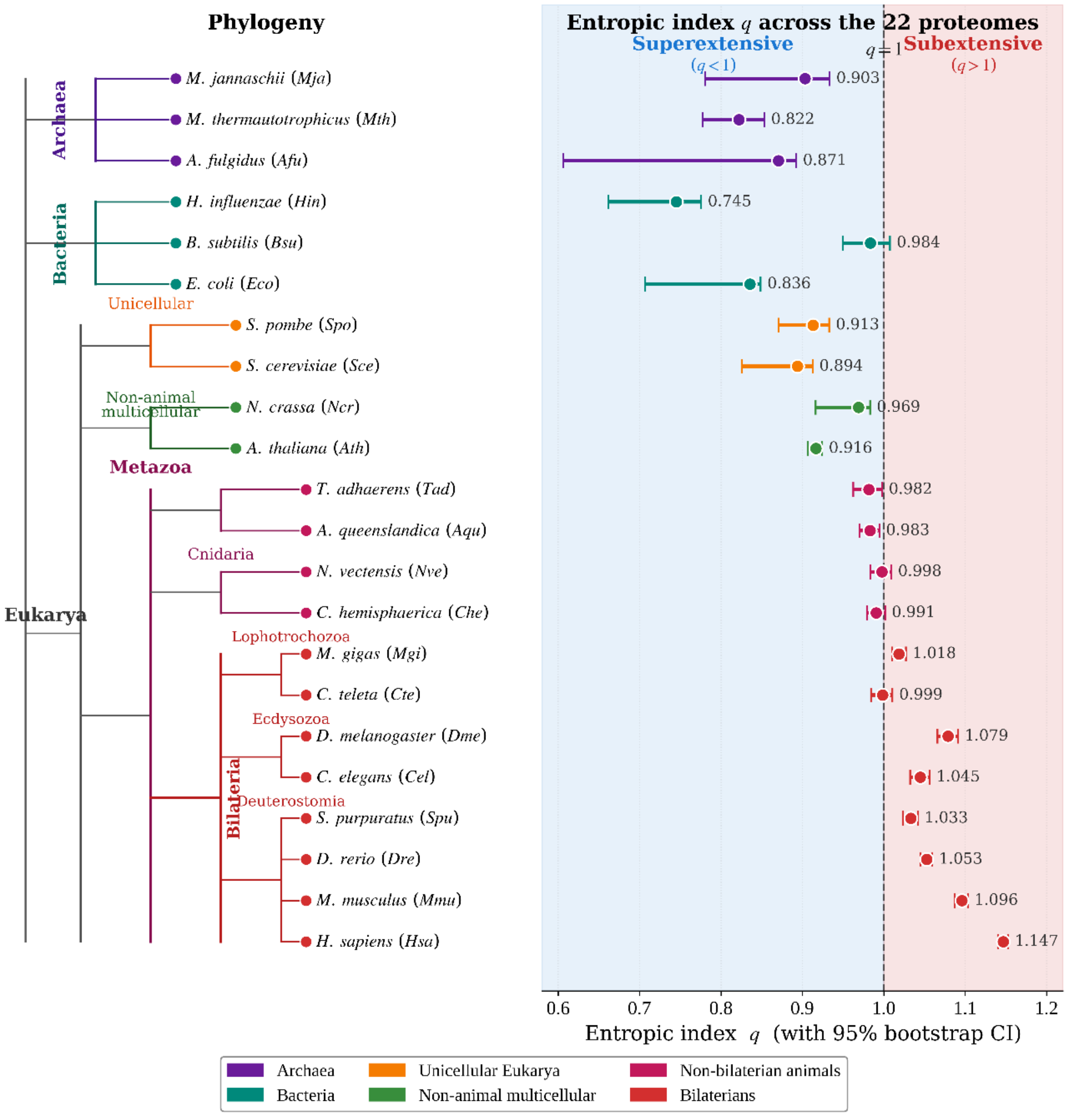

# Figure 2

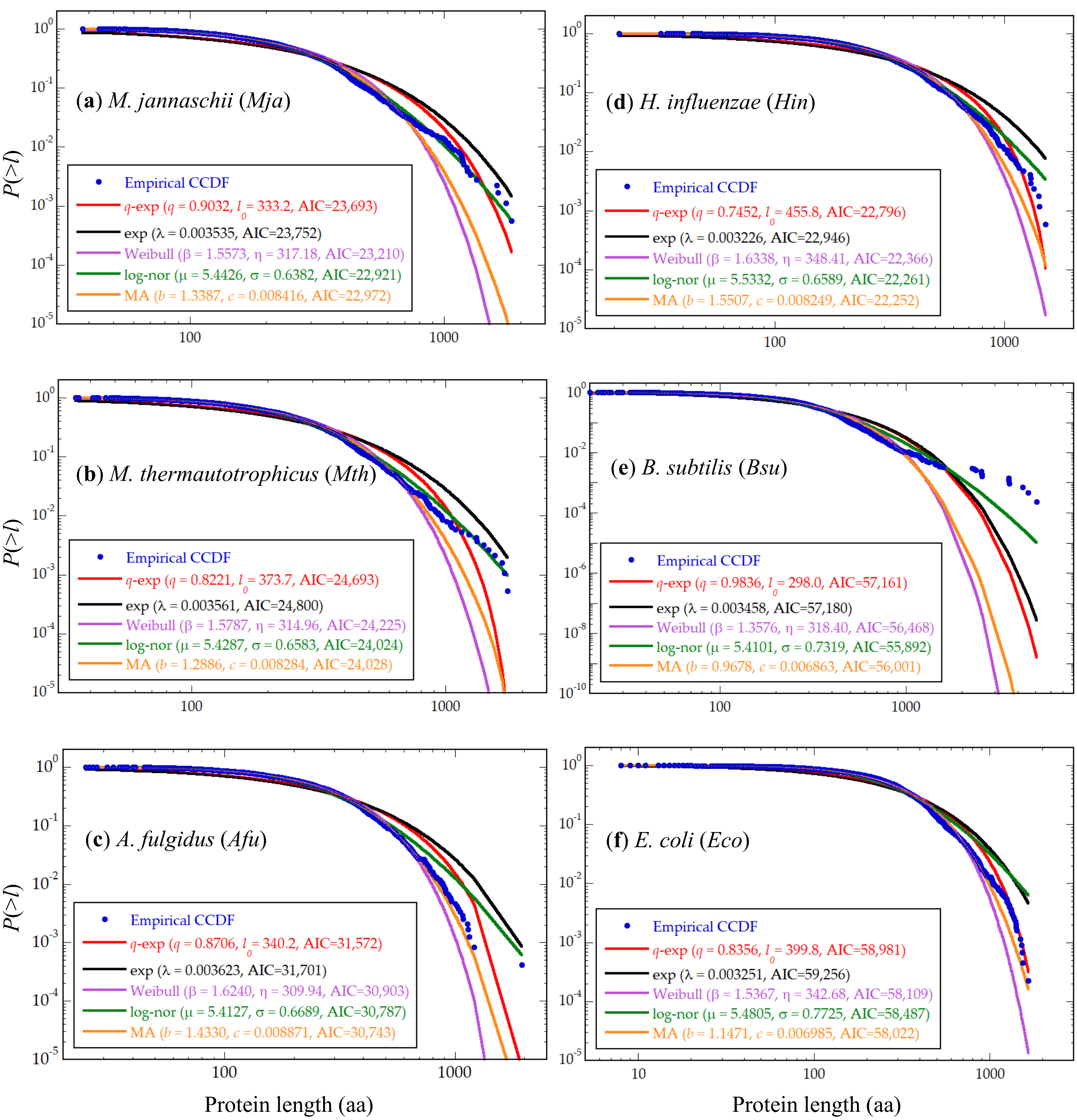

(a) M. jannaschii (Mja)
Empirical CCDF
q-exp (q = 0.9032, l0 = 333.2, AIC=23,693)
exp (λ = 0.003535, AIC=23,752)
Weibull (β = 1.5573, η = 317.18, AIC=23,210)
log-nor (μ = 5.4426, σ = 0.6382, AIC=22,921)
MA (b = 1.3387, c = 0.008416, AIC=22,972)
(d) H. influenzae (Hin)
Empirical CCDF
q-exp (q = 0.7452, l0 = 455.8, AIC=22,796)
exp (λ = 0.003226, AIC=22,946)
Weibull (β = 1.6338, η = 348.41, AIC=22,366)
log-nor (μ = 5.5332, σ = 0.6589, AIC=22,261)
MA (b = 1.5507, c = 0.008249, AIC=22,252)
(b) M. thermautotrophicus (Mth)
Empirical CCDF
q-exp (q = 0.8221, l0 = 373.7, AIC=24,693)
exp (λ = 0.003561, AIC=24,800)
Weibull (β = 1.5787, η = 314.96, AIC=24,225)
log-nor (μ = 5.4287, σ = 0.6583, AIC=24,024)
MA (b = 1.2886, c = 0.008284, AIC=24,028)
(e) B. subtilis (Bsu)
Empirical CCDF
q-exp (q = 0.9836, l0 = 298.0, AIC=57,161)
exp (λ = 0.003458, AIC=57,180)
Weibull (β = 1.3576, η = 318.40, AIC=56,468)
log-nor (μ = 5.4101, σ = 0.7319, AIC=55,892)
MA (b = 0.9678, c = 0.006863, AIC=56,001)
(c) A. fulgidus (Afu)
Empirical CCDF
q-exp (q = 0.8706, l0 = 340.2, AIC=31,572)
exp (λ = 0.003623, AIC=31,701)
Weibull (β = 1.6240, η = 309.94, AIC=30,903)
log-nor (μ = 5.4127, σ = 0.6689, AIC=30,787)
MA (b = 1.4330, c = 0.008871, AIC=30,743)
(f) E. coli (Eco)
Empirical CCDF
q-exp (q = 0.8356, l0 = 399.8, AIC=58,981)
exp (λ = 0.003251, AIC=59,256)
Weibull (β = 1.5367, η = 342.68, AIC=58,109)
log-nor (μ = 5.4805, σ = 0.7725, AIC=58,487)
MA (b = 1.1471, c = 0.006985, AIC=58,022)
P(>l)
Protein length (aa)

**Figure 2** (continued)

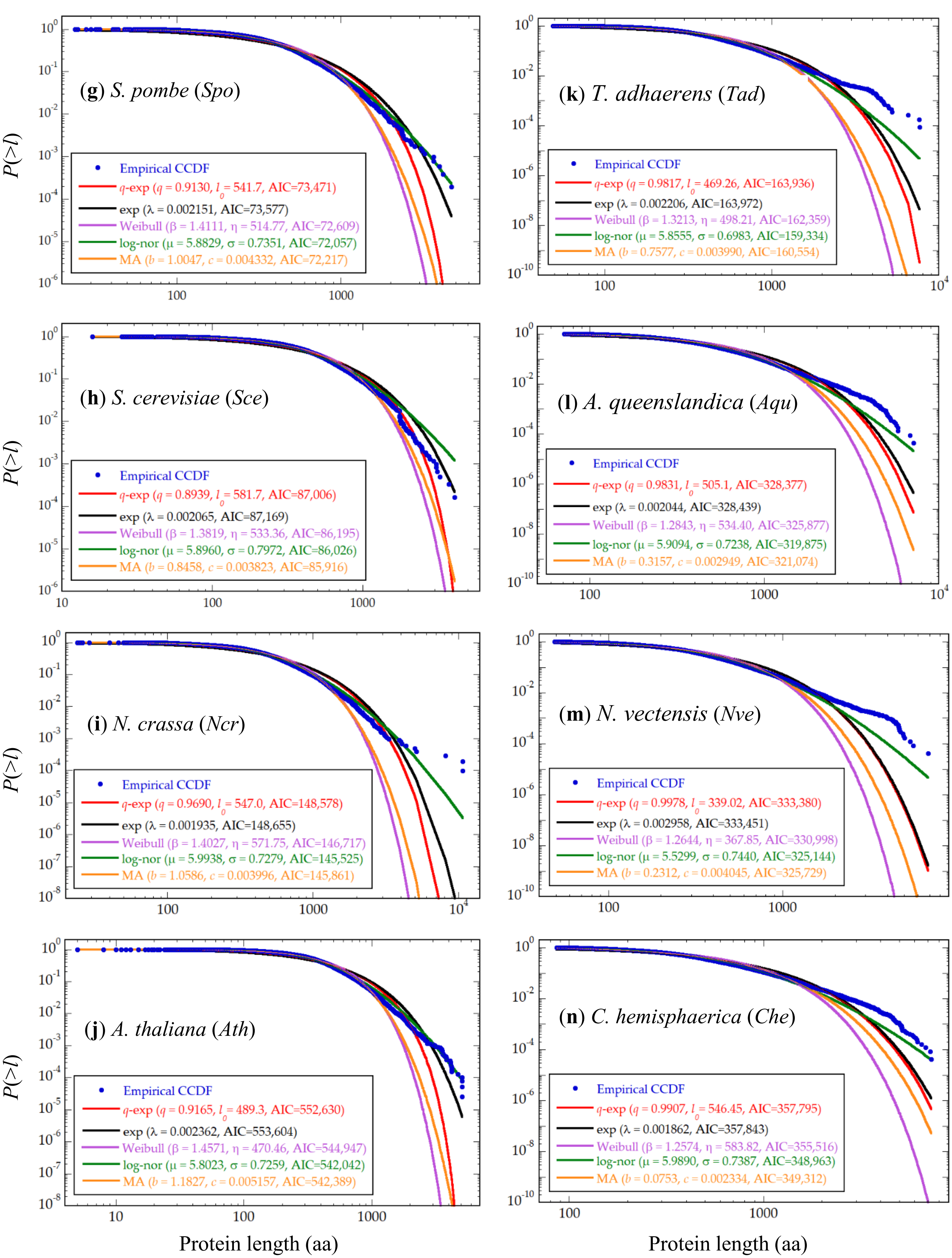

**Figure 2** (continued)

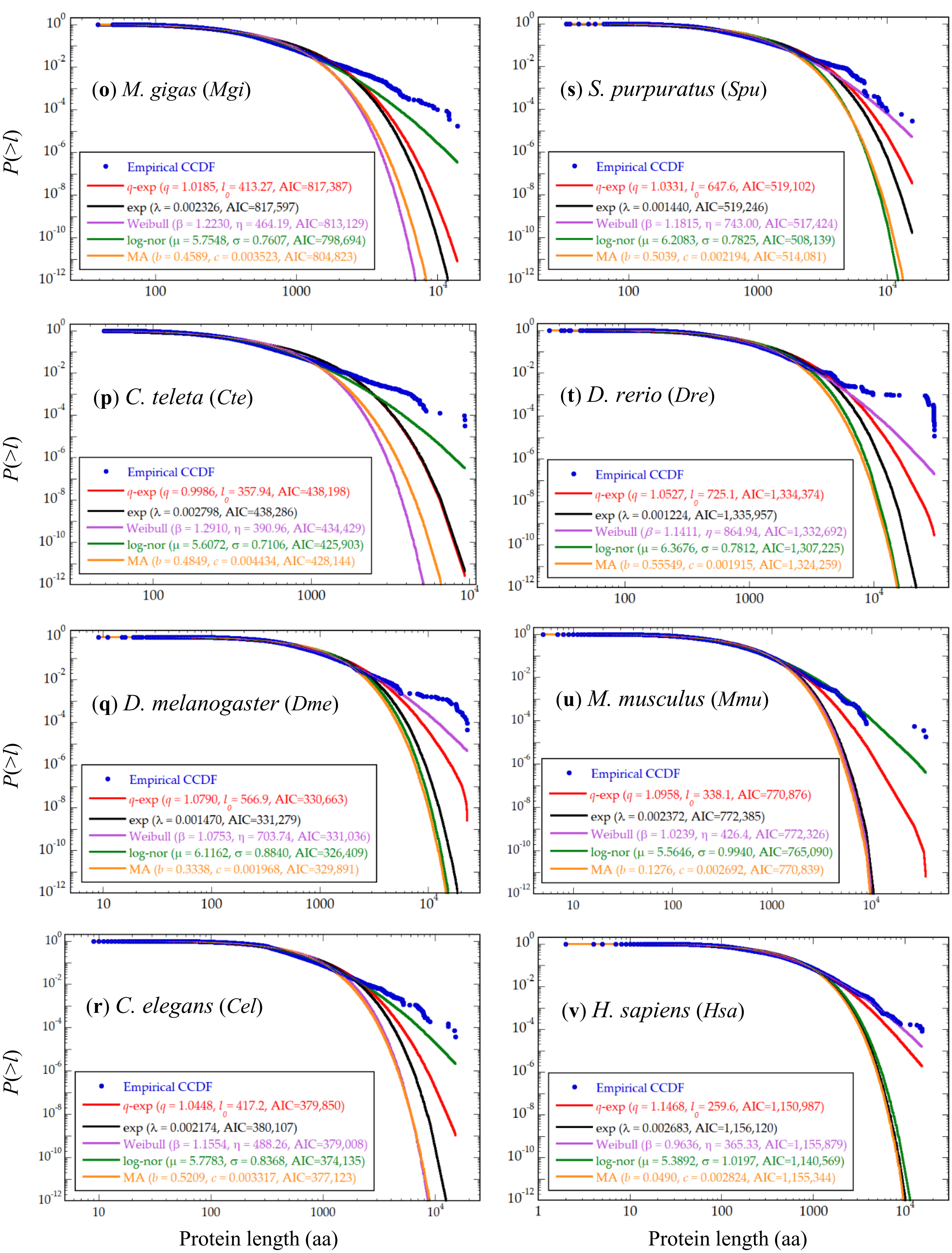

# Figure 3

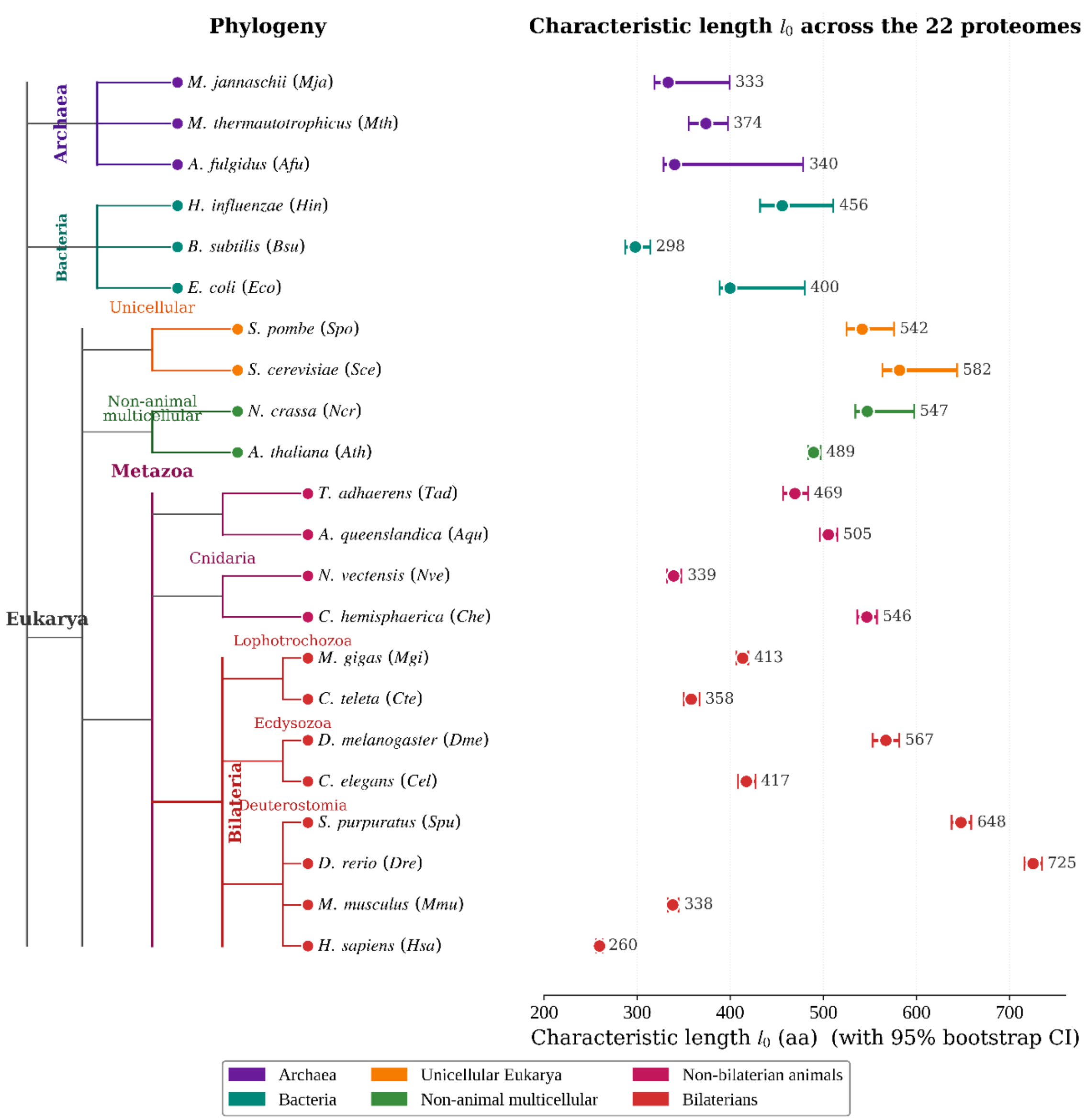

Phylogeny
Characteristic length $l_0$ across the 22 proteomes
Archaea
Bacteria
Eukarya
Unicellular
Non-animal multicellular
Metazoa
Cnidaria
Lophotrochozoa
Ecdysozoa
Deuterostomia
Bilateria
M. jannaschii (Mja)
M. thermautotrophicus (Mth)
A. fulgidus (Afu)
H. influenzae (Hin)
B. subtilis (Bsu)
E. coli (Eco)
S. pombe (Spo)
S. cerevisiae (Sce)
N. crassa (Ncr)
A. thaliana (Ath)
T. adhaerens (Tad)
A. queenslandica (Aqu)
N. vectensis (Nve)
C. hemisphaerica (Che)
M. gigas (Mgi)
C. teleta (Cte)
D. melanogaster (Dme)
C. elegans (Cel)
S. purpuratus (Spu)
D. rerio (Dre)
M. musculus (Mmu)
H. sapiens (Hsa)
333
374
340
456
298
400
542
582
547
489
469
505
339
546
413
358
567
417
648
725
338
260
200
300
400
500
600
700
Characteristic length $l_0$ (aa) (with 95% bootstrap CI)
Archaea
Bacteria
Unicellular Eukarya
Non-animal multicellular
Non-bilaterian animals
Bilaterians